\documentclass[twocolumn,letterpaper,aps,prc,superscriptaddress,showpacs,floatfix,preprintnumbers,longbibliography]{revtex4-1}

\usepackage{graphicx}   
\usepackage{xspace}     
\usepackage{url}
\usepackage[mathlines]{lineno}
\usepackage{bm}
\usepackage{amsmath}
\usepackage{xcolor}
\usepackage{comment}
\usepackage[driverfallback=dvipdfm]{hyperref}
\hypersetup{
	colorlinks = true,
	citecolor= blue,
	linkbordercolor = cyan
}

\newcommand{\mean}[1]{\langle #1 \rangle}
\newcommand{\dmean}[1]{\langle \langle \, #1\, \rangle\rangle}

\def \be {\begin{equation}}
\def \ee {\end{equation}} 
\def \bea {\begin{eqnarray}}
\def \eea {\end{eqnarray}}

\def \psirp {\Psi_{\rm RP}}
\def \psin {\Psi_{n}}
\def \psina {\Psi_{n}^a}
\def \psinb {\Psi_{n}^b}
\def \psinc {\Psi_{n}^c}
\def \psind {\Psi_{n}^d}
\def \Res {\rm Res}

\usepackage{color}
\definecolor{orange}{cmyk}{0.,0.353,1.,0.}    

\usepackage{lineno}
\newcommand*\patchAmsMathEnvironmentForLineno[1]{
  \expandafter\let\csname old#1\expandafter\endcsname\csname #1\endcsname
  \expandafter\let\csname oldend#1\expandafter\endcsname\csname end#1\endcsname
  \renewenvironment{#1}
     {\linenomath\csname old#1\endcsname}
     {\csname oldend#1\endcsname\endlinenomath}}
\newcommand*\patchBothAmsMathEnvironmentsForLineno[1]{
  \patchAmsMathEnvironmentForLineno{#1}
  \patchAmsMathEnvironmentForLineno{#1*}}
\AtBeginDocument{
\patchBothAmsMathEnvironmentsForLineno{equation}
\patchBothAmsMathEnvironmentsForLineno{align}
\patchBothAmsMathEnvironmentsForLineno{flalign}
\patchBothAmsMathEnvironmentsForLineno{alignat}
\patchBothAmsMathEnvironmentsForLineno{gather}
\patchBothAmsMathEnvironmentsForLineno{multline}
}

\begin{document}


\title{Measurement of the longitudinal flow-plane decorrelation using multi-plane cumulants in \texorpdfstring{$\sqrt{s_{_{\mathrm{NN}}}}$}{snn}\xspace~=~200~GeV Au+Au, Ru+Ru, and Zr+Zr collisions}

\author{The STAR Collaboration}

\begin{abstract} 

Measurements of the variation of anisotropic flow-plane angles ($\Psi_n$) with rapidity, commonly known as the flow-plane decorrelation, provide important insights into the initial conditions of the matter produced in heavy-ion collisions. 
In this paper, using data collected by the STAR experiment, we report the first measurement of the four-plane correlator observable $T_{n}\{ba;dc\}=\langle\langle\sin [n(\Psi^{b}_{n}-\Psi^{a}_{n})]\sin[n(\Psi^{d}_{n}-\Psi^{c}_{n})]\rangle\rangle$, where superscripts $a$, $b$, $c$, and $d$ denote sequential pseudorapidity ($\eta$) regions with $a$ corresponding to the most backward region, $b$ and $c$ close to mid-rapidity with $\eta_b<0$ and $\eta_c>0$, and $d$ being the most forward. 
The measurement is performed for the elliptic and triangular flow (i.e.~$n=2$ and $3$) in Au+Au and isobar (Ru+Ru, Zr+Zr) collisions at $\sqrt{s_{_{\mathrm{NN}}}}$\xspace~=~200~GeV. 
The goal of calculating the correlation of the flow-plane angle variations from backward to mid-central, and from mid-central to forward regions, is to probe the systematic variation of flow angle over a wide $\eta$ range. 
In mid-central collisions ($10-30\%$ centrality), we find $T_{2}\{ba;dc\}= -0.004\pm 0.001 (\rm stat)\pm0.002(\rm syst)$ independent of the collision system. 
Such a small value of $T_{2}$ favors a ``random-walk'' variation of the flow-plane angles, where the rapidity correlation length is smaller than the entire region under study. 
These measurements provide new information on the decorrelation patterns in the system and offer a quantitative estimate of possible systematic variations in anisotropic flow angles such as ``twist'' between forward and backward regions. 
This opens new opportunities for understanding the three-dimensional structure and the time evolution of the quark-gluon plasma created in heavy-ion collisions.
\end{abstract}


\maketitle

\setlength\linenumbersep{0.10cm}

\section{Introduction}
\label{intro}

The STAR experiment at the Relativistic Heavy Ion Collider (RHIC) at Brookhaven National Laboratory (BNL) focuses on investigating the formation of a deconfined state of quarks and gluons known as the quark-gluon plasma (QGP)~\cite{ADAMS2005102,Busza:2018rrf}. 
One of the primary goals of STAR and other heavy ion experiments is to probe the mechanism behind the QGP formation and to understand its evolution and hadronization processes. 
It has long been realized that the QGP created in such collisions exhibits significant initial-state spatial anisotropies which originate from both the geometry of non-central nuclear collisions and the probabilistic nature of nucleon-nucleon interactions. 
The spatial anisotropies in the initial-state matter distribution lead to momentum anisotropies of the final-state particles due to the space-momentum correlations during system evolution.

The azimuthal anisotropy in the distribution of emitted particles can be characterized by a superposition of different modes or harmonics of collective flow in a Fourier series~\cite{ZPhysC.70.1671,  PhysRevC.58.1671, Voloshin:2008dg}
\begin{equation}
  \label{eq:anisotropy}
    \frac{dN}{d\phi}\propto 1 + 2\sum_{n}^{\infty} v_{n}\cos[n(\phi-\Psi_{n})],
 \end{equation}
where $\phi$ is the azimuthal emission angle of the particle, $v_n$ characterizes the strength of the $n$-th harmonic flow, and $\psin$ is the flow (symmetry) plane angle.  
Flow coefficients $v_{1}$, $v_{2}$, and $v_{3}$ are often referred to as ``directed", ``elliptic", and ``triangular" flow, respectively.

The reaction plane angle $\psirp$ is defined as a plane spanned by the impact parameter vector and the beam axis. Using $\psirp$ in place of $\psin$ in the decomposition Eq.~\ref{eq:anisotropy} for all harmonics, corresponds to the flow ($v_n$) definition in the reaction-plane anisotropic flow picture. 
While this picture is well defined theoretically, in practice it is not very useful, as the reaction-plane angle can not be measured experimentally. 
In most cases, the flow itself (e.g.,~anisotropic particle distribution) is used to determine the flow-symmetry planes $\Psi_n$ as those corresponding to the maximum amplitude of a given harmonic~\cite{Voloshin:2008dg}. 
Due to the initial-state geometry fluctuations, flow planes $\psin$ in general do not coincide with the reaction plane~\cite{Alver:2010gr}.  
In this so-called participant picture of flow, the flow development is primarily driven by the initial geometry of the system, including fluctuations caused by the probabilistic character of the interactions of nucleons inside the colliding nuclei. 
The participant-plane description is meaningful only if the flow angles depend relatively weakly on the particle rapidity and, more generally, on the particle transverse momentum ($p_\mathrm{T}$) as well as particle type. 
Otherwise, the Fourier decomposition in Eq.~\ref{eq:anisotropy} is not practically useful.

The study of the flow-plane angle $\psin$ variation in longitudinal direction, along with its fluctuations and correlations in $\eta$, is of particular interest. 
Together with similar measurements of the flow coefficients $v_n$ 
they provide insights into various aspects of nuclear interactions, such as baryon stopping, shadowing, and sub-nucleon structure.

For discussion purposes, it is useful to separate the different sources of flow-plane decorrelations. Short-range effects, such as jet energy deposition, influence only a limited region in pseudorapidity. 
Long-range effects, on the other hand, arise from correlations like those in the ``twisted fireball'' scenario~\cite{PhysRevC.83.034911,PhysRevResearch.2.023362}, where the flow plane in the forward region is primarily determined by the participant distribution from the projectile nucleus, while at backward pseudorapidity, it is governed by the participants of target nucleus. The short-range effects when measured on a distance scale much larger than the correlation length appear as a ``random-walk'' variations in the flow angle. 
The resulting decorrelation is referred to as a random tilt. 
The ``twisted fireball'' scenario, on the other hand, can be visualized as a monotonic change in the flow-plane angle with pseudorapidity. 
Another scenario corresponding to the long-range variation of flow plane occurs when the flow-plane angles in the forward and backward regions are close to each other, while those at the central region experiences relatively large fluctuations. 
This case would correspond to a non-monotonic change in the flow-plane angle with pseudorapidity. 
See Refs.~\cite{PhysRevC.91.044904,cite-key, WU2019327,PhysRevC.90.034905, PhysRevC.90.034915, Xiao:2012uw,Jia:2024xvl} for a more detailed discussion on the mechanisms of the flow-plane decorrelations.

\begin{figure}[htbp!]
    \begin{center}
    \includegraphics[width =\columnwidth]{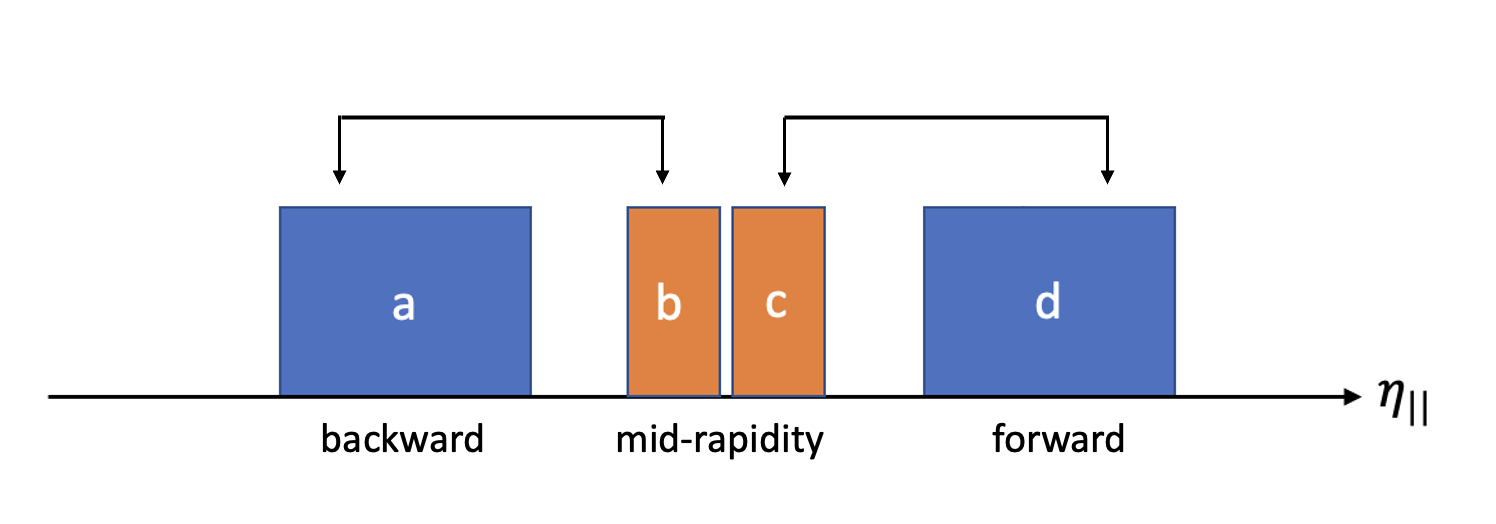}
    \caption{Diagram illustrating four pseudorapidity regions used  
    for $T_{n}$ calculations.      }
    \label{fig:t2-fig}
    \end{center}
\end{figure}

Over the past decade, to study the flow decorrelation, different collaborations~\cite{PhysRevC.92.034911,  EurPhysJC.78.142, 
PhysRevLett.126.122301} 
often used an observable known as the factorization ratio:
\begin{equation}\label{eq:ratio}
\begin{split}
  r_{n}(\eta_{a\rightarrow c/b}) & \equiv
  \frac{\langle\cos[n(\phi^{c}-\Psi_{n}^{a})]\rangle}
       {\langle\cos[n(\phi^{b}-\Psi_{n}^{a})]\rangle}
       \\&=
       \frac{\langle
         v_{n}(\eta_{c})\cos[n(\Psi_{n}^{c}-\Psi_{n}^{a})]\rangle}
            {\langle
              v_{n}(\eta_{b})\cos[n(\Psi_{n}^{b}-\Psi_{n}^{a})]\rangle}
\end{split}
\end{equation}
where $\langle ...\rangle$ denotes the average over events. $\phi^{b}$ and $\phi^{c}$ are the azimuthal angles of particles from the two symmetric mid-pseudorapidity windows $b$ and $c$, respectively, on the backward and forward regions (see Fig.~\ref{fig:t2-fig}). 
Similarly, $\Psi_{n}^{a}$, $\Psi_{n}^{b}$, and $\Psi_{n}^{c}$ are the flow-plane angles of $n^{\rm th}$-order harmonic flow obtained from the reference $\eta$ window $a$ ($\eta_a<0$), and two symmetric  windows $b$ ($\eta_{b}<0$) and $c$ ($\eta_{c}>0$). 
The notations $\eta_{a}$, $\eta_{b}$, and $\eta_{c}$ refer to the ranges in pseudorapidity of three windows such that the reference window $a$ is sufficiently far from windows $b$ and $c$ to introduce large pseudorapidity gaps, $|\eta_{b}-\eta_{a}|$ and $|\eta_{c}-\eta_{a}|$, in the measurement, ensuring suppression of non-flow. 
The reference window in Eq.~\ref{eq:ratio} can also be selected from the forward region (such as window $d$ in Fig.~\ref{fig:t2-fig}), provided that regions $b$ and $c$ are simultaneously interchanged in the definition.

In symmetric collisions, such as Au+Au or Ru+Ru (Zr+Zr), the flow coefficients $v_{n}(\eta_{b})$ and $v_{n}(\eta_{c})$, corresponding to symmetric negative and positive ranges $\eta_b$ and $\eta_c$, respectively, are expected to be identical when averaged over all events. 
Hence, the deviation of the ratio in Eq.~\ref{eq:ratio} from unity would indicate the presence of flow decorrelation in the collisions. 
The observable $r_{n}(\eta_{a\rightarrow c/b})$ in Eq.~\ref{eq:ratio} includes contributions from the decorrelations in the flow angles as well as in the flow magnitudes.
As shown in Ref.~\cite{Bozek:2017qir}, the decorrelations in the flow magnitudes are relatively small, at the level of a few percent. 
However, the effect of using the flow magnitude weight in the calculation of the flow-plane decorrelation can be rather significant and comparable to the flow-angle decorrelation itself. 
This is due to the fact that the flow-plane fluctuations are smaller in the events with larger flow magnitude~\cite{Bozek:2017qir}.
Experimentally, the flow (event) planes are determined from flow itself. This determination is more accurate in events with larger flow, and thus the measurements of flow-plane decorrelations always implicitly include flow magnitude weights. Below, in the discussion of the results of our measurements in terms of flow-plane decorrelations we assume that the corresponding average values are obtained with flow magnitude weights. 

The observable $r_{n}(\eta_{a\rightarrow c/b})$ is sensitive to both the random tilt and the long-range systematic rotation of flow planes as in the ``twisted fireball'' scenario.
As shown in Ref.~\cite{Xu:2020koy}, the ratio $r_{n}(\eta_{a\rightarrow c/b})$ is independent of the reference pseudorapidity window $a$ in the random-walk scenario of the flow-plane decorrelations.

In the small angle  approximation, the observable $r_{n}(\eta_{a\rightarrow c/b})$ can be expressed as 
\begin{equation}\label{ratio-expn}
\begin{split}
    &r_{n}(\eta_{a\rightarrow c/b})\\ &\approx 1-\frac{n^{2}}{2}\langle(\Delta\Psi_{n}^{b\rightarrow c})^{2}\rangle-n^{2}\langle\Delta\Psi_{n}^{a\rightarrow b}\cdot\Delta\Psi_{n}^{b\rightarrow c}\rangle,
\end{split}
\end{equation}
using the relationships
\begin{equation}
\begin{split}
    \Delta\Psi_{n}^{a\rightarrow c} &= \Delta\Psi_{n}^{a\rightarrow b} + \Delta\Psi_{n}^{b\rightarrow c},\\
    \Delta\Psi_{n}^{a\rightarrow c} &=\Psi_{n}^{c}-\Psi_{n}^{a},\\
    \Delta\Psi_{n}^{a\rightarrow b} &=\Psi_{n}^{b}-\Psi_{n}^{a},\\
    \Delta\Psi_{n}^{b\rightarrow c} &= \Psi_{n}^{c}-\Psi_{n}^{b},
\end{split}
\end{equation}
where $\Delta\Psi_{n}^{a\rightarrow c}$, $\Delta\Psi_{n}^{a\rightarrow b}$, and $\Delta\Psi_{n}^{b\rightarrow c}$ represent the flow-plane twists, describing the variation of $\Psi_{n}$ between the regions $a\rightarrow c$, $a\rightarrow b$, and $b\rightarrow c$, respectively.

The observable defined in Eq.~\ref{eq:ratio} is designed to measure the relative decrease in the anisotropic flow caused by the $\Psi_{n}$ decorrelation over a $\eta_{c}-\eta_{b}$ distance. 
In practice, the decorrelation measured by Eq.~\ref{eq:ratio} is independent of the reference window $a$ only if there is no contribution from the term $\langle\Delta\Psi_{n}^{a\rightarrow b}\cdot\Delta\Psi_{n}^{b\rightarrow c}\rangle$ (see Eq.~\ref{ratio-expn}), i.e.,~if the consecutive changes in the flow-plane orientations  $\Delta\Psi_{n}^{a\rightarrow b}$ and $\Delta\Psi_{n}^{b\rightarrow c}$ of the flow-plane angles are independent.

This paper presents a study of the longitudinal flow-plane decorrelations using a four-plane $T_n$ correlator, proposed in Ref.~\cite{Xu:2020koy}, using the flow-plane angles from the backward, forward, and two mid-rapidity regions.
The $T_n$ observable is defined by 
\begin{eqnarray}
\label{T2def}
 T_n\{ba;dc\}  
    &\equiv&
    \dmean{\sin [n(\psinb-\psina)]  \sin [n(\psind-\psinc)]} 
    \\&=&
    \mean{\sin [n(\psinb-\psina)]\sin [n(\psind-\psinc)]}
    \label{eq:sinsin}
    \\&&-
    \frac{1}{2}\mean{\cos [n(\psinb-\psind)]} \mean{\cos [n(\psina-\psinc)]}
  \nonumber 
  \\&&+
  \frac{1}{2}\mean{\cos [n(\psina-\psind)]}  \mean{\cos [n(\psinb-\psinc)]}, 
\nonumber 
\end{eqnarray}
where double angle brackets $\langle\langle ...\rangle\rangle$ denote the cumulant part, obtained by subtracting from the 4-plane correlator (first line in Eq.~\ref{eq:sinsin}) the contributions from all lower-order correlations (subsequent lines). 
$\Psi^{i}_{n} \textrm{ (for } i=a,b,c,d)$ are the flow-plane angles of $n^{\rm th}$-order harmonic flow. 
The superscript in the flow-plane angle denotes subevents, $a$ through $d$, as illustrated in Fig.~\ref{fig:t2-fig}. 
In the expansion Eq.~\ref{eq:sinsin} we do not show the terms that are present in the full cumulant expansion of the four-plane correlator but that are equal to zero due to azimuthal symmetry (e.g. proportional to averages of sine or cosine of a single flow-plane angle; see Ref.~\cite{Xu:2020koy} for details). 
In the analysis of the real data, to account for small anisotropies in particle detection the vanishing of such terms is achieved by {\em recentering} of the flow vectors, see below. 

The main property of the observable $T_n$ is that it is nonzero only if all four flow-plane angles are correlated.
Thus it is largely insensitive to correlations unrelated to the flow-plane orientation (i.e., non-flow correlations due to few-particle clusters such as jets, heavy resonance decays, local charge conservation,~etc.), which are suppressed similarly to that in flow measurements based on fourth-order cumulants~\cite{PhysRevC.64.054901,Voloshin:2008dg,PhysRevC.66.034904}.
The AMPT studies of $T_n$ observable~\cite{Xu:2020koy} confirm this conclusion. 
For the same reason, $T_n$ is also insensitive to the ``random-walk'' type of flow decorrelations. 
Similarly to the ratio observable $r_n(\eta_{a\rightarrow c/b})$, the exprimental measurements of $T_n$ are sensitive to the flow magnitude fluctuations and  effectively include flow magnitudes as a weight.

The flow-plane decorrelations are expected to be relatively small. In this case,
\begin{equation}
  T_{n}\{ba;dc\}
    \approx \dmean{n\Delta\Psi^{a\rightarrow
      b}_{n} \cdot n\Delta\Psi^{c\rightarrow d}_{n}},
\end{equation}
where $\Delta\Psi^{a\rightarrow
b}_{n}=\Psi_{n}^{b}-\Psi_{n}^{a}$ and $\Delta\Psi^{c\rightarrow d}_{n}=\Psi_{n}^{d}-\Psi_{n}^{c}$.
Therefore, $T_n$ represents the covariance between changes in orientation of flow plane from window $a$ to $b$ and that from $c$ to $d$, quantifying the overall azimuthal twist of flow plane on moving forward in the pseudorapidity. 
One would expect a positive value of $T_{n}\{ba;dc\}$ when the flow-plane angles at backward and forward regions are fluctuating in opposite directions to that at mid-rapidity.  
On the other hand, the negative sign of $T_{n}\{ba;dc\}$ would correspond to the decorrelation pattern when the flow-plane angles at backward and forward regions fall on the same direction relative to that at the central region. 
These two distinct cases $T_{n}>0$ and $T_{n}<0$ are identified as the ``S-shaped" and ``C-shaped" decorrelations, respectively, in Ref.~\cite{Xu:2020koy}.

In this analysis, we attempt to determine and identify the dominant pattern of the longitudinal flow-plane decorrelations by measuring $r_n$
and $T_n$ observables using the flow-plane angles reconstructed in the forward, backward, and mid-pseudorapidity regions in isobar (Ru+Ru, Zr+Zr) and Au+Au collisions at $\sqrt{s_{_\mathrm{NN}}}=200$~GeV with the STAR detector at RHIC.

\section{EXPERIMENT AND DATA ANALYSIS}

\subsection{Datasets and Event Selection}
\begin{figure*}[htbp!]
\par\medskip
\centerline{
\includegraphics[angle=0,width=1.0\columnwidth]{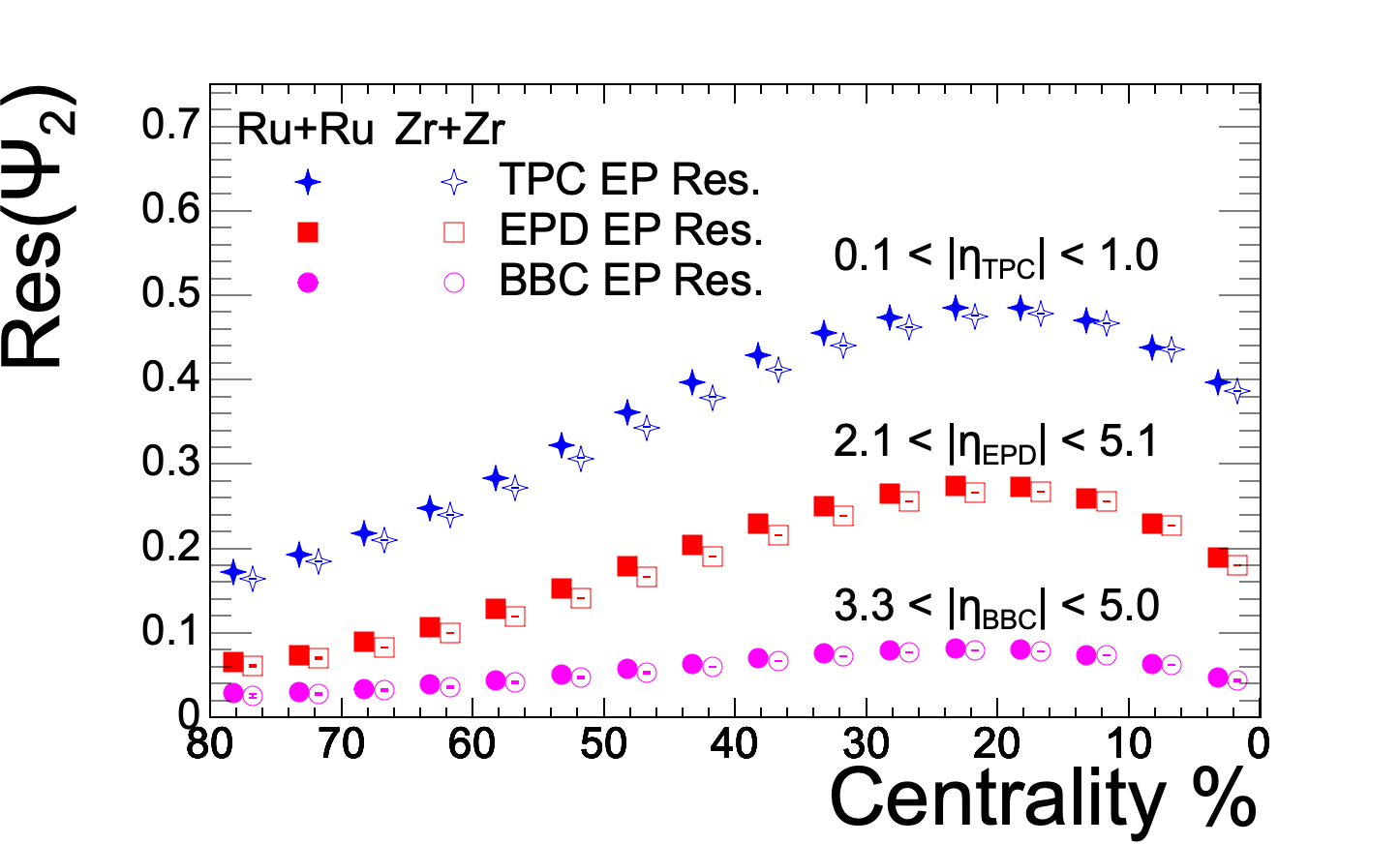}
\includegraphics[width=1.0\columnwidth]{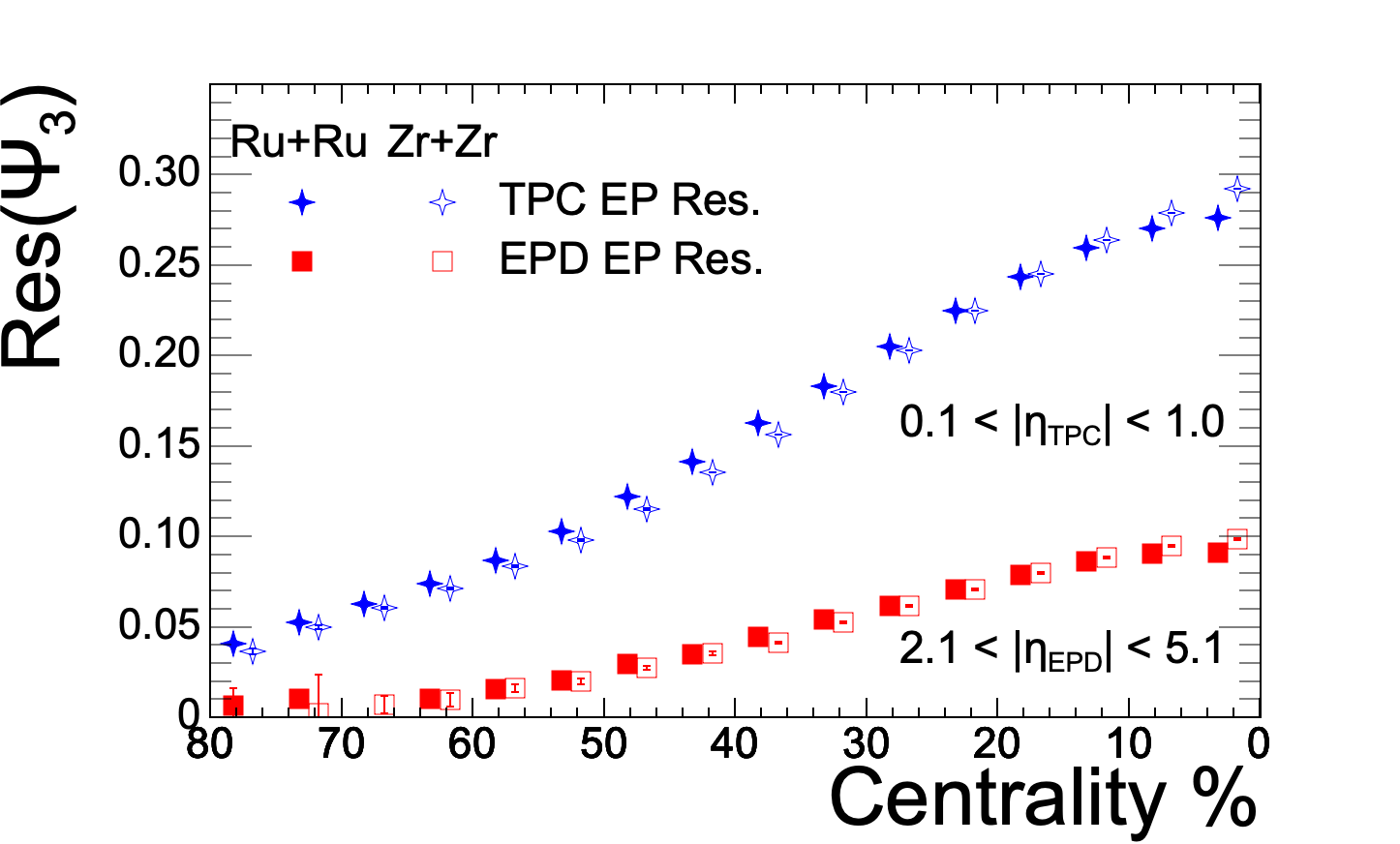}}
\caption{  Left: The second harmonic event plane resolutions from the TPC,  EPD, and BBC detectors. 
  Right: The third  harmonic sub-event plane resolutions from the TPC and EPD. 
  The filled and open data points are for Ru$+$Ru and Zr$+$Zr collisions, respectively. 
  The Zr$+$Zr data points are shifted towards right along the x-axis for clarity. 
  The results shown are from the two-subevent method.}
\label{fig:psi2psi3Res}
\end{figure*}

The primary dataset used in this work was collected by the STAR experiment~\cite{HARRIS1994277} for isobar Ru+Ru and Zr+Zr collisions at $\sqrt{s_{_\mathrm{NN}}}=200$~GeV in $2018$ to search for the Chiral Magnetic Effect (CME)~\cite{PhysRevC.105.014901}. 
The main detectors used in the analysis are the Time Projection Chamber (TPC)~\cite{ANDERSON2003659}, Event Plane Detector (EPD)~\cite{ADAMS2020163970}, and Beam-Beam Counter (BBC)~\cite{10.1063/1.2888113}. 
To study the system size dependence of the $T_{n}$ observable, we also analyze the data of Au$+$Au collisions at $\sqrt{s_{_\mathrm{NN}}}=200$~GeV taken in years $2011$ and $2016$. The EPD subsystem used in the analysis of the isobar collisions is an detector upgrade installed in the STAR detector system in $2018$. 
It significantly improves the measurements of the event-plane angle in forward and backward pseudorapidity regions, compared to the BBC detector used earlier (see Fig.~\ref{fig:psi2psi3Res}).
Since the Au+Au collisions data in $2011$ and $2016$ were taken using the TPC and BBC, the BBC is used in the analyses of these datasets to estimate the event-plane orientations at forward and backward pseudorapidities.
The results using Au+Au and isobar collisions are then corrected by corresponding BBC or EPD event-plane resolutions.

Many details of this analysis are similar to those in the CME isobar analysis. 
The event and the track selections, as described in details in Ref.~\cite{PhysRevC.105.014901}, have been used on the same datasets for this analysis. The vertex distribution  on isobar collisions exhibited asymmetry with a peak around~$5$~cm due to online vertex selection. Therefore, an asymmetric vertex position cut of $-35 <v_{z} < 25$~cm on the primary $z-$vertex measured using a coordinate system with the origin at the TPC center was applied to maximize the statistics. Additional cut of $v_{r}  =\sqrt{v_{x}^{2}+v_{y}^{2}}  <  2$~cm in the transverse plane was also applied for each collision event. 
In order to ensure the selection of good events, requirement of $|v_{z,  \mathrm{TPC}} -v_{z, \mathrm{VPD}}| <5$~cm was imposed on the primary collision vertices, where $v_{z, \mathrm{VPD}}$ represents the primary interaction position along the beam direction measured by the Vertex Position Detectors (VPDs)~\cite{LLOPE2004252}. 
In addition, we require tracks reconstructed in the TPC to have a distance of closest approach (DCA) to the primary vertex to be less than $3$ cm. 
We also require each track to have greater than $15$ ionization  points reconstructed in the TPC. 

The symmetric vertex position cuts of $-30 <v_{z} < 30$~cm and $-6 <v_{z} < 6$~cm were applied to the $2011$ and $2016$ Au+Au collision datasets, respectively, as online vertex reconstruction for these collisions indicated symmetric $v_{z}$ distributions. Due to the deployment of the Heavy Flavor Tracker (HFT)~\cite{Bouchet:2009id} as an upgrade to the inner tracking system of the STAR experiment, a relatively narrow $v_{z}$ cut was applied for the $2016$ Au+Au collision dataset. Other track selection cuts used for the Au+Au collision analysis are same as those used for the isobar collision analysis.

\subsection{Event Planes Reconstruction}

The second- and third-order flow vectors $Q_{n}\equiv(Q_{n,x},  Q_{n,y})$ and the {\em event-plane} angles $\psi_{n}$ (an estimate of the flow-plane angle $\Psi_n$) are reconstructed from the anisotropic flow itself following the procedure described in Refs.~\cite{PhysRevC.58.1671,  Voloshin:2008dg}:
\begin{eqnarray}\label{flow-vectors}
    Q_{n,x} = \sum w_{i}\cos(n\phi_{i}),\label{Qx}\\
    Q_{n,y} = \sum w_{i}\sin(n\phi_{i}),\label{Qy}\\
    \psi_{n} = \frac{1}{n}\arctan\left(\frac{Q_{n,y}}{Q_{n,x}}\right),
\end{eqnarray}
where $w_{i}$ is the weight for $i^{th}$ particle, and the sum goes over all particles used in the calculation. 
For the flow vector and event-plane angle reconstruction in the TPC, primary tracks with $0.2< p_{\mathrm{T}} < 2.0$~GeV/$c$ were taken from $-1 < \eta < -0.1$ and $0.1 <\eta < 1$.
The flow vectors were reconstructed with unit weight for the Particle-level calculations and with transverse momentum weight for the Q-level calculations (see below). 
Charged particles emitted in the forward and backward pseudorapidity regions create a signal distribution with distinct peaks, each representing different numbers of minimally ionizing particles (nMIP) in the EPD and BBC tiles. 
In the EPD, the event-plane angles were reconstructed using nMIP of each EPD tile $i$ which satisfies $0.3\leq$ nMIP $\leq 3$ as weight, $w_{i}$, in Eqs.~\ref{Qx} and ~\ref{Qy}. 
The weight is set to $3$ if nMIP $> 3$ to suppress distortions from large Landau fluctuations in the energy-loss distribution. 
Similar information is used to reconstruct event-plane angles in the BBC.

Recentering of flow vectors and flattening of $\psi_{n}$ distribution were performed to remove any acceptance effects. 
This procedure was done on the run-by-run basis in each z-vertex bin of $2$~cm width and centrality bin of $5\%$ step.

Figure~\ref{fig:psi2psi3Res} shows the event-plane resolutions 
${\rm Res}(\Psi_n)\equiv\mean{\cos[n(\psi_n-\Psi_n)]}$ as a function of centrality for the second- and third-harmonic anisotropic flow calculated for the TPC, EPD, and BBC pseudorapidity subevents in the isobar collisions. 
The event-plane resolution calculations were done with the two- and three-subevent methods~\cite{Voloshin:2008dg}. Figure~\ref{fig:psi2psi3Res} presents calculations using the two-subevent method.
Note that the calculation of the event-plane resolutions using rapidity subevents might be affected by flow-plane decorrelations. See the next section for more details. 

\begin{figure*}[htbp!]
\centerline{\includegraphics[angle=0, width=1.0\columnwidth]{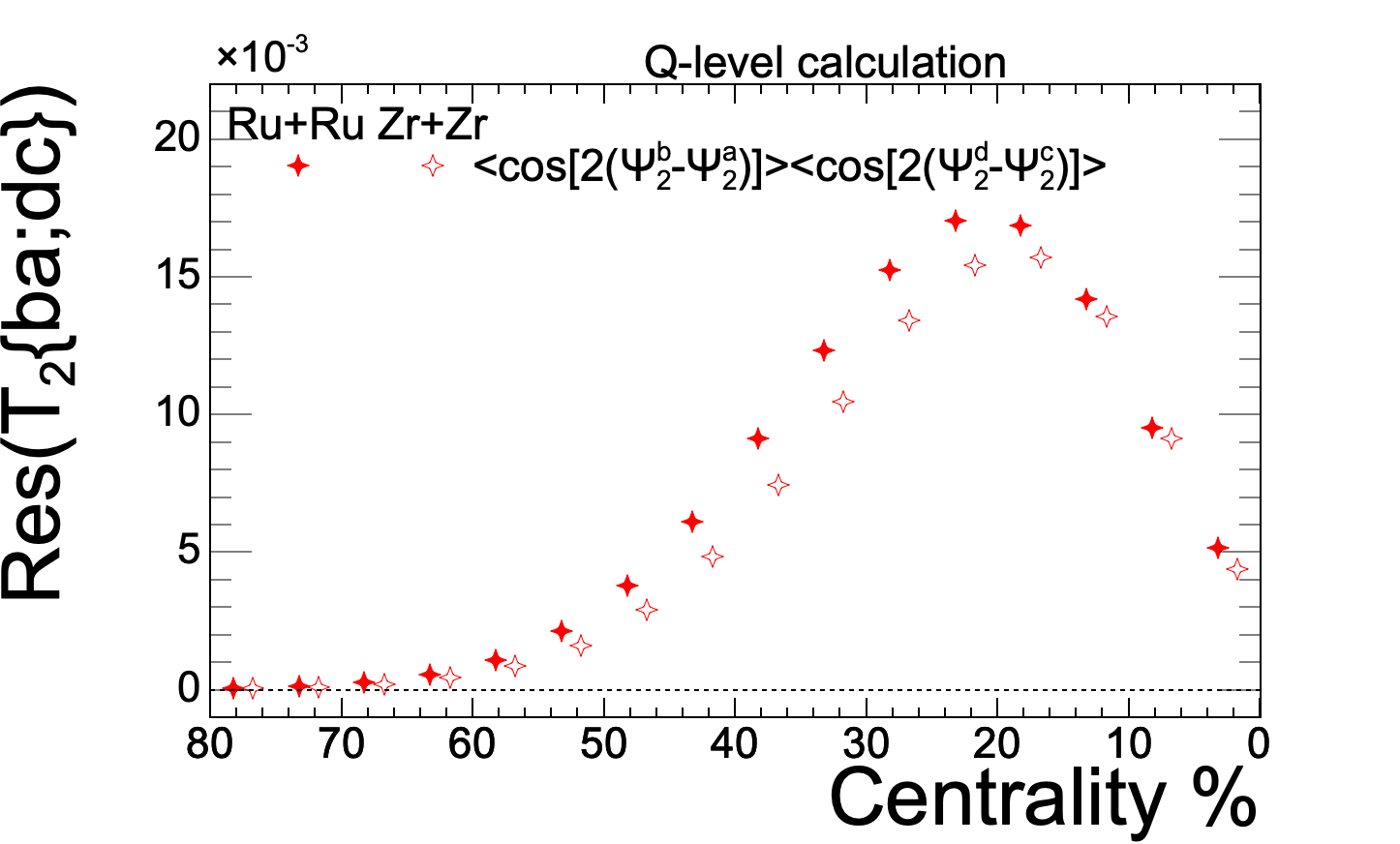}\includegraphics[width=1.0\columnwidth]{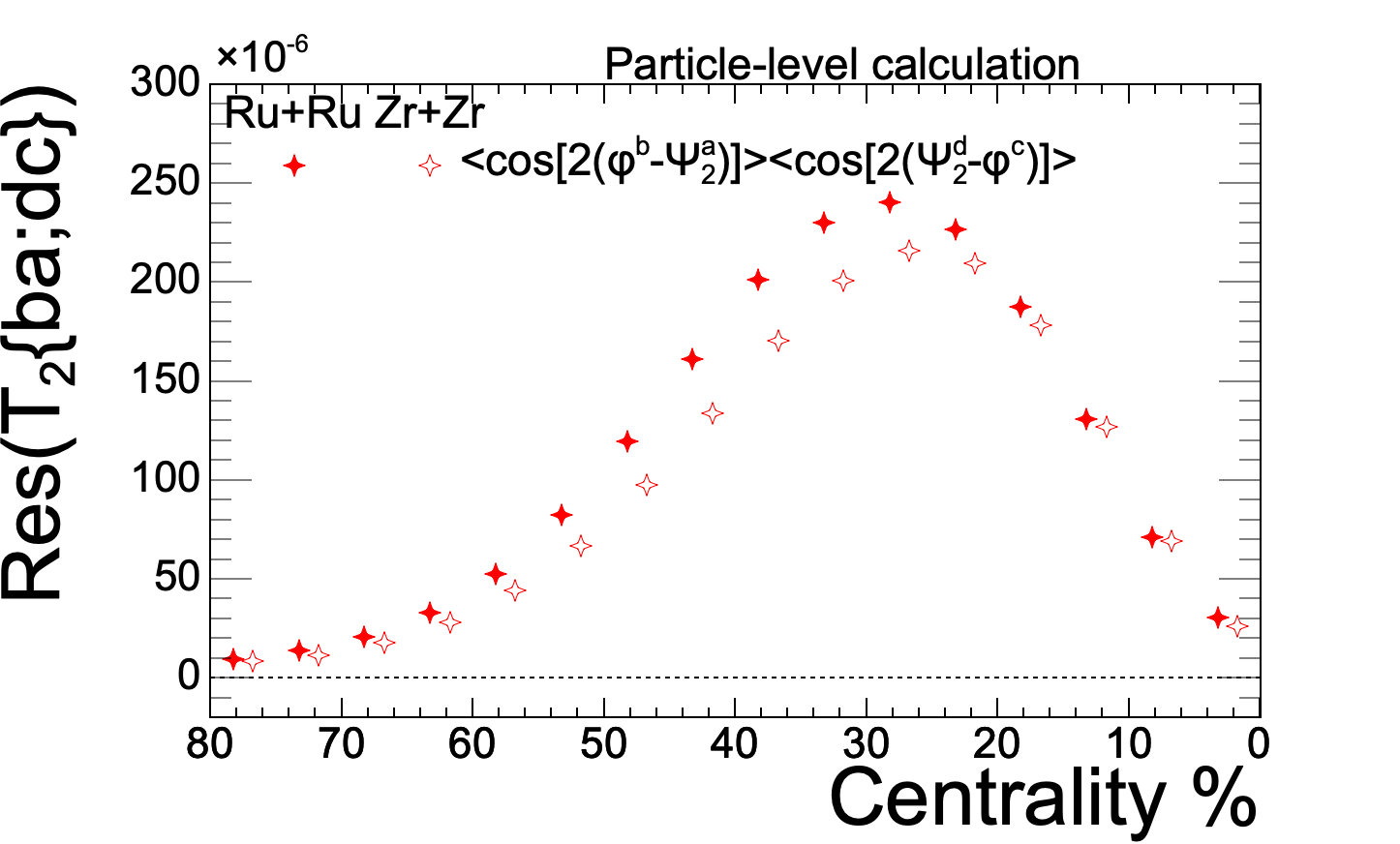}}
\caption{(Color online) Resolution ${\Res}(T_{n}\{ba;dc\})$ plots for the 
  Q-level (left) and Particle-level (right) calculations in Ru$+$Ru (solid markers) and Zr$+$Zr (open markers) collisions for the second-order anisotropic flow. 
  The Zr$+$Zr data points are slightly shifted along x-axis for
  clarity.}
\label{fig:t2badc-isobar-res}
\end{figure*}

\begin{figure*}[htbp!]
\centerline{\includegraphics[angle=0,
    width=1.0\columnwidth]{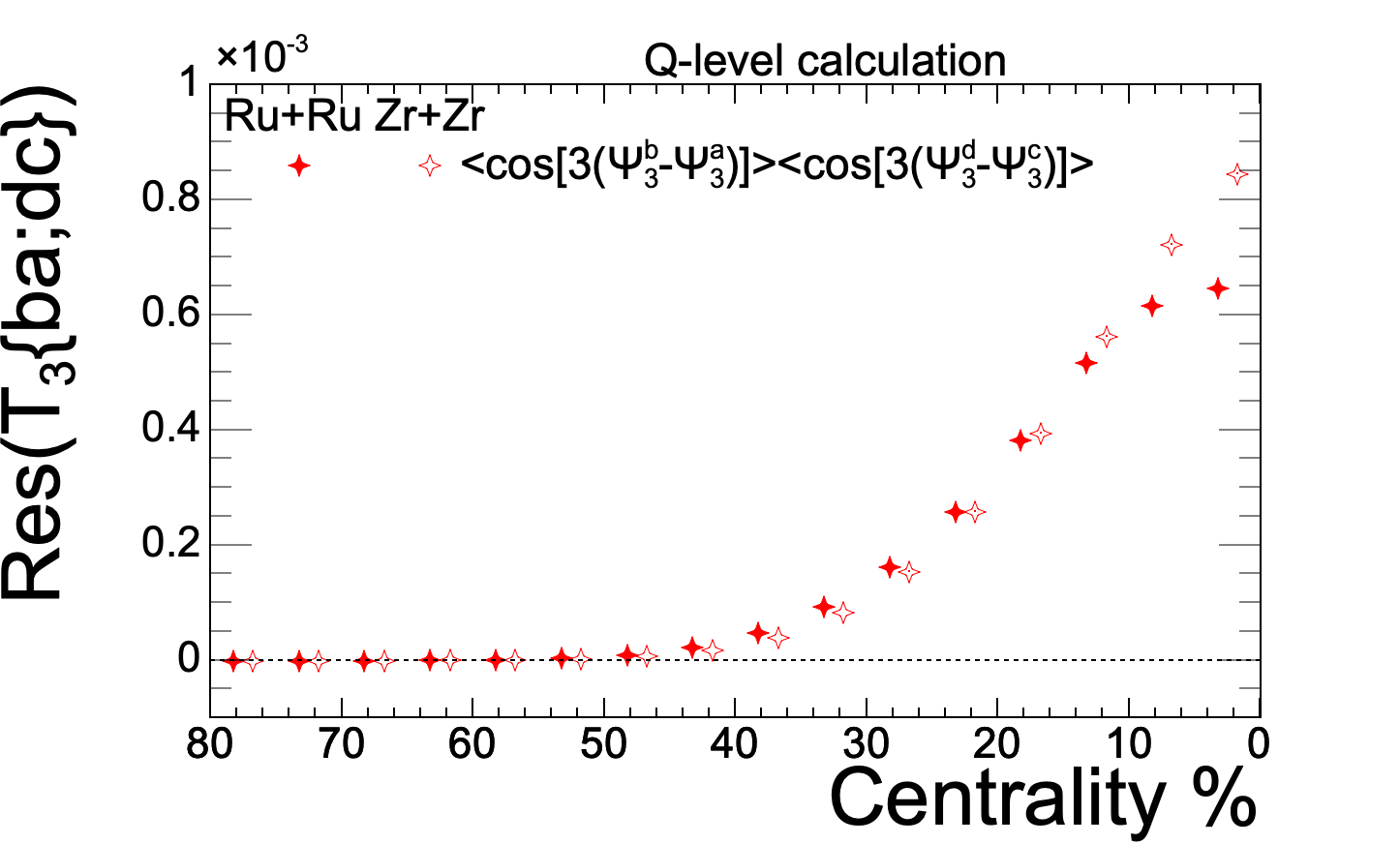}\includegraphics[width=1.0\columnwidth]{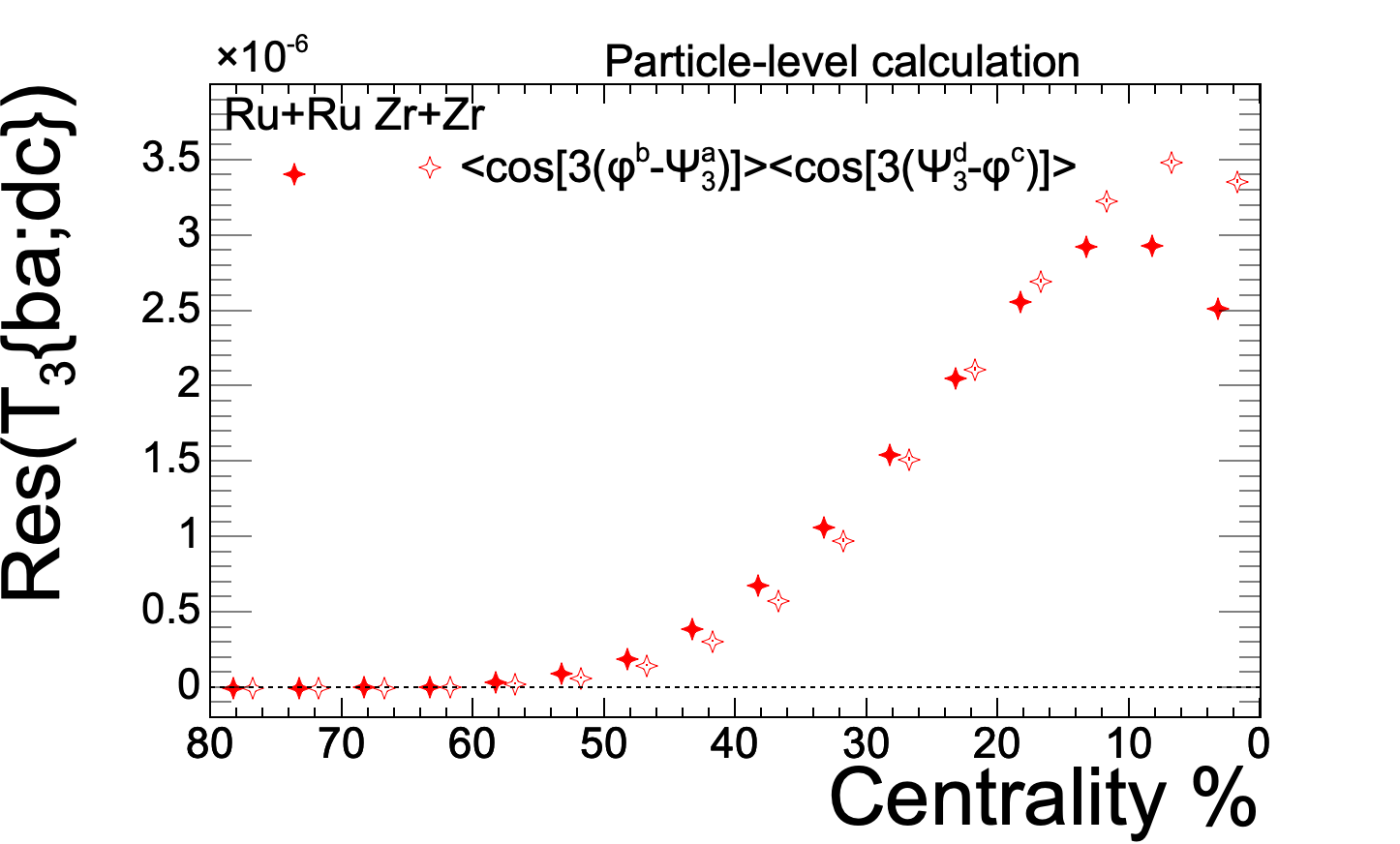}}
\caption{(Color online) Same as in  Fig.~\ref{fig:t2badc-isobar-res} for the third-order  anisotropic flow.
}
\label{fig:t3badc-isobar-res}
\end{figure*}
\subsection{Calculation of the Four-Plane Correlator}
\label{intro-tn}

In this analysis, the measurements of the $T_{n}\{ba;dc\}$ correlators for the second- and third-order anisotropic flow are performed using event-plane angles, $\psi_n$'s, reconstructed in the STAR TPC and EPD (or BBC) detectors.
The pseudorapidity windows $a$, $b$, $c$, and $d$, depicted in Fig.~\ref{fig:t2-fig}, correspond to the EPD-east ($-5.1 < \eta < -2.1$) or BBC-east ($-5.0 < \eta < -3.3$), TPC-east ($-1.0 < \eta < -0.1$), TPC-west ($0.1 < \eta < 1.0$),  and EPD-west ($2.1 < \eta < 5.1$) or BBC-west ($3.3 < \eta < 5.0$) regions, respectively.

The measurement using four event planes, from the east and west regions of the TPC and EPD (or BBC), is referred to as the {Q-level} calculation, 
and the measurement using two event planes from the east and west regions of the EPD (or BBC) and the particle's azimuth $\phi$ from the east/west regions of the TPC is identified as the {Particle-level} calculation. 

\begin{figure*}[htbp!]
\centerline{\includegraphics[angle=0,
    width=1.0\columnwidth]{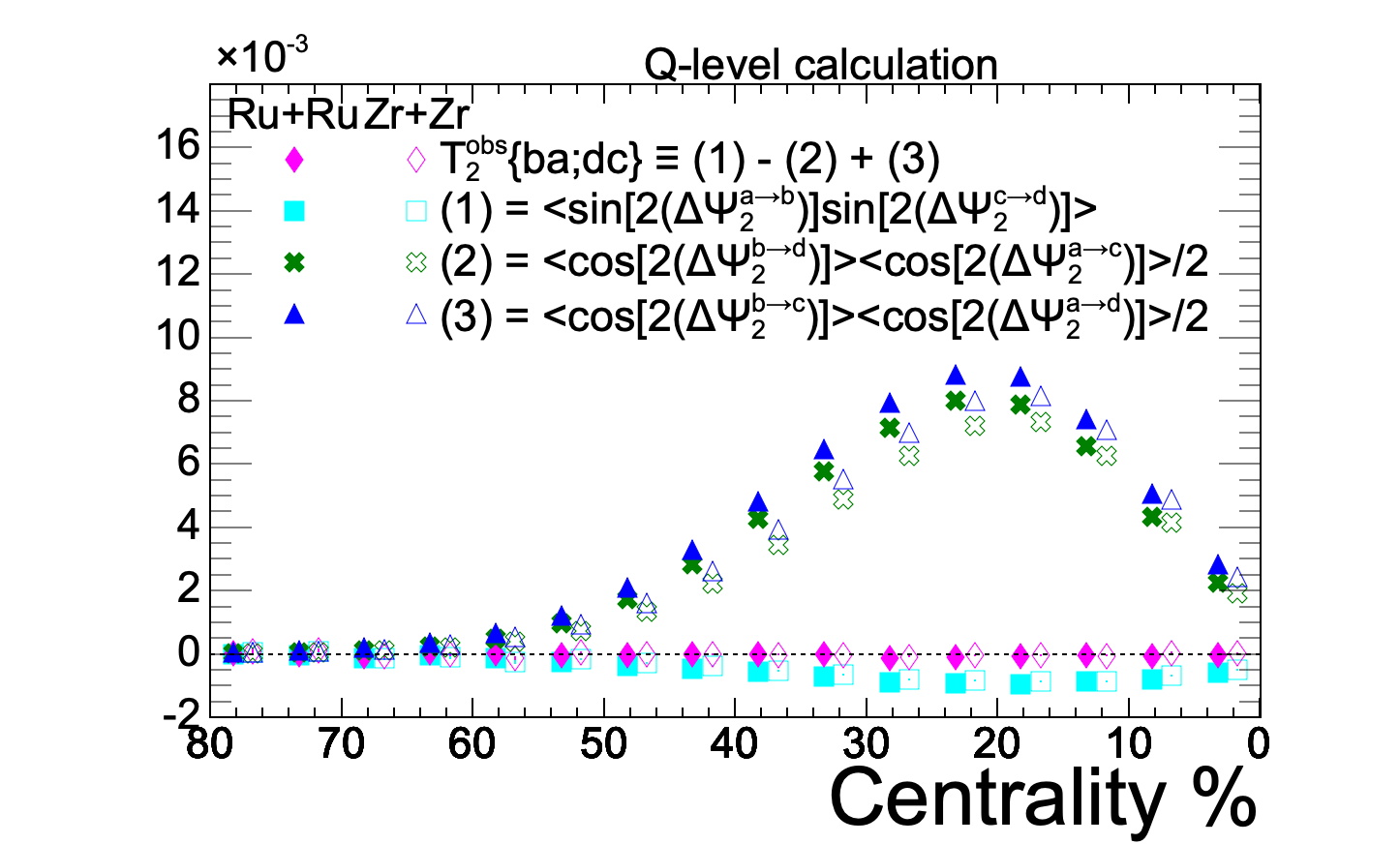}\includegraphics[width=1.0\columnwidth]{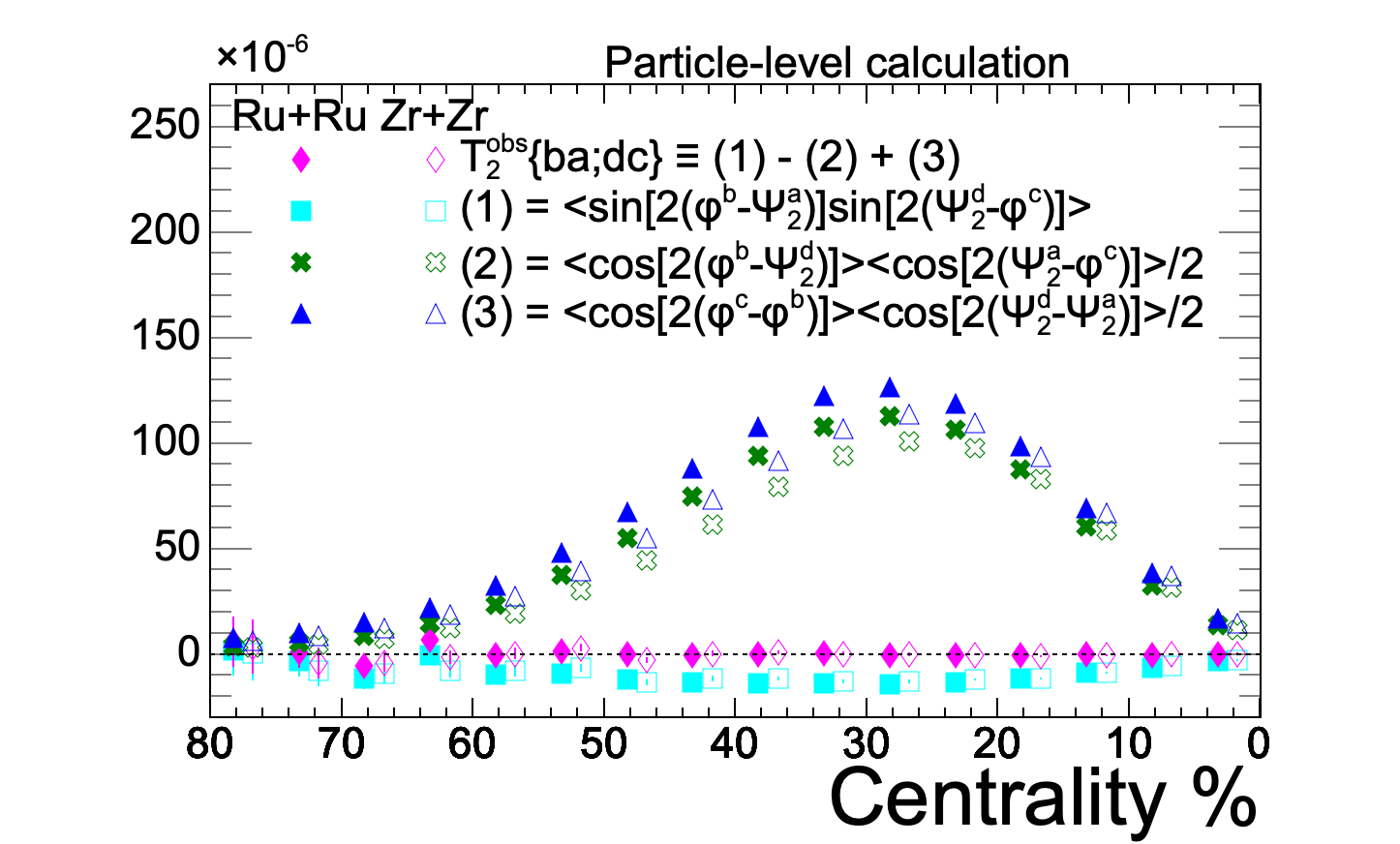}} 
\caption{(Color online) Centrality dependence of different terms contributing to
  $T^{obs}_{2}\{ba;dc\}$, as given in  Eqs.~\ref{T2explict} and \ref{T2explictpl}:
  Solid markers show Ru$+$Ru and open markers Zr$+$Zr results for the 
  Q-level (left) and Particle-level (right) approaches. 
     The Zr$+$Zr data points are shifted towards right for clarity. 
}
\label{fig:t2badc-isobar-all-comp}
\end{figure*}
\begin{figure*}[htbp!]
\centerline{\includegraphics[angle=0,
    width=1.0\columnwidth]{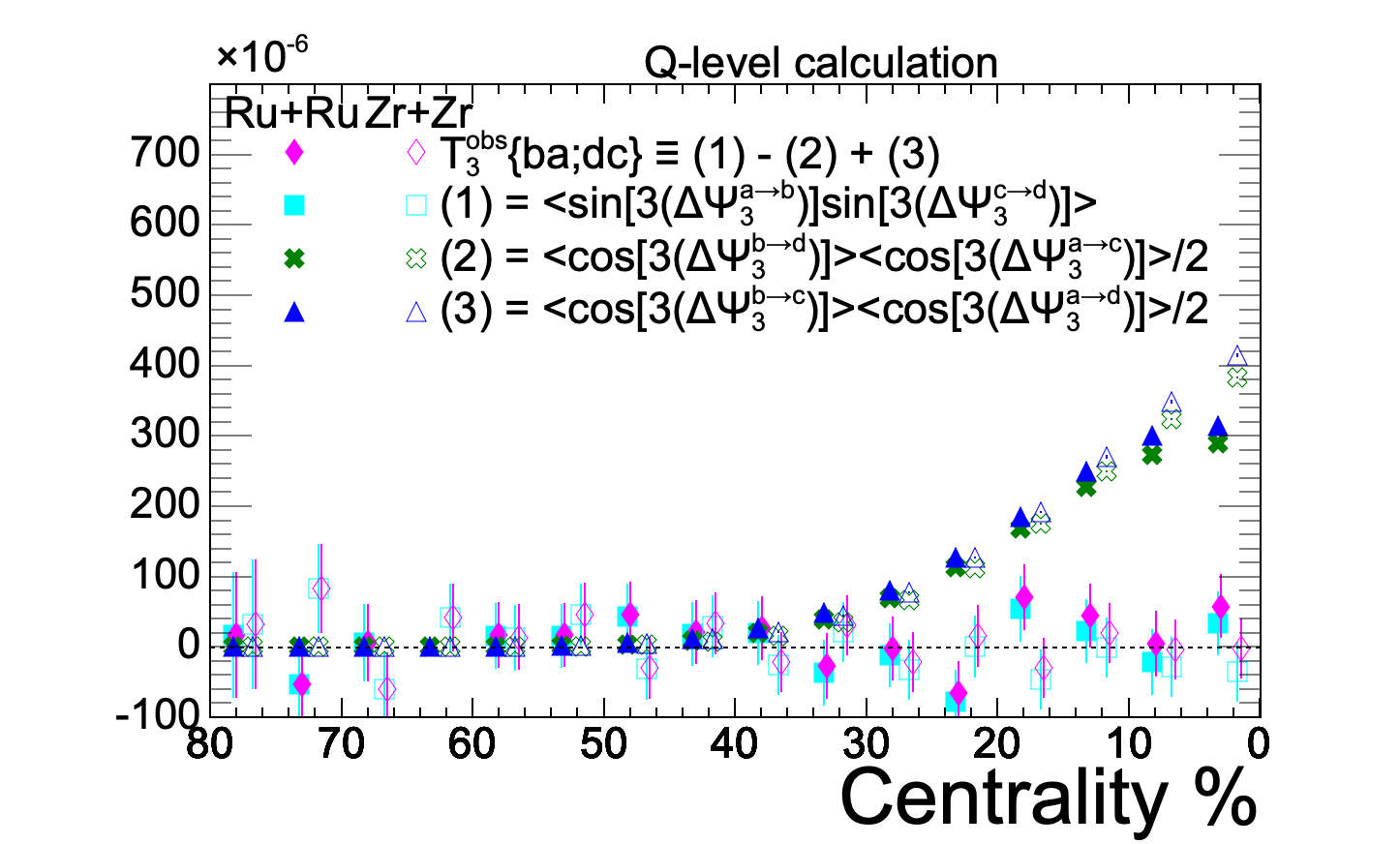}\includegraphics[width=1.0\columnwidth]{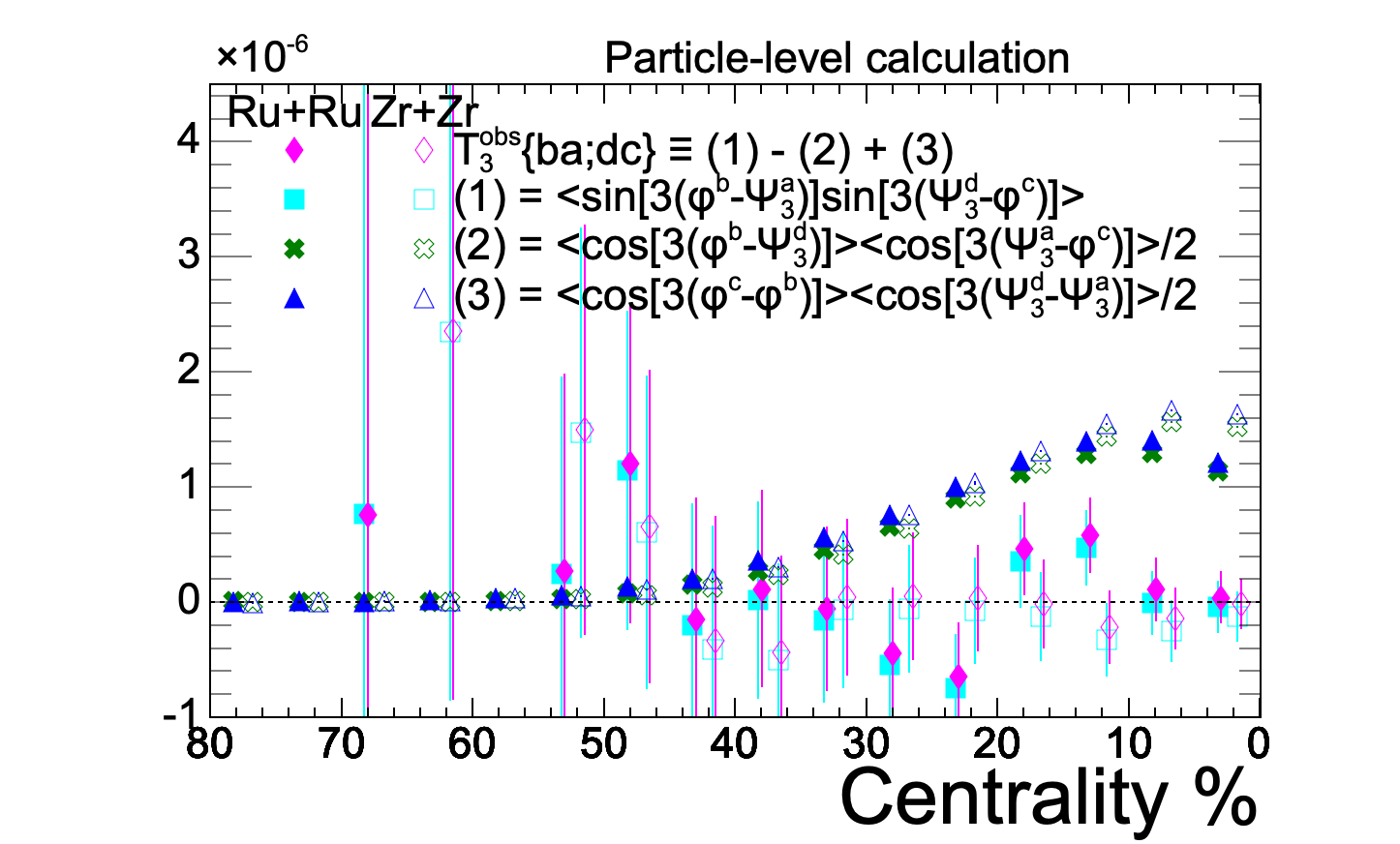}}
\caption{Same as 
  Fig.~\ref{fig:t2badc-isobar-all-comp} for the third-order
  anisotropic flow $n=3$.
  }
\label{fig:t3badc-isobar-all-comp}
\end{figure*}
%
%
Experimentally, the four-plane correlator  $T_{n}$ is calculated as
\begin{equation}
  \label{Tn}
  T_{n}\{ba;dc\}(\mathrm{Q\,\text{-}\,level})
  =\frac{T_{n}^{obs}\{ba;dc\}}{\Res{\it (T_{n}\{ba;dc\})}},
\end{equation}
where the term ${\Res}(T_{n}\{ba;dc\})$ in the denominator is used to correct $T^{obs}_{n}\{ba;dc\}$ for the event-plane resolution, and discussed below.
The numerator, the ``observed'' $T_n^{obs}$ correlator, for Q-level calculations is
evaluated following the $T_n$ definition, Eq.~\ref{T2def}:
\begin{equation}\label{T2explict}
\begin{split}
  &T^{obs}_{n}\{ba;dc\}
  \\&\equiv
  \langle\langle\sin [n(\psi^{b}_{n}-\psi^{a}_{n})]
  \sin [n(\psi^{d}_{n}-\psi^{c}_{n})]\rangle\rangle
  \\&=
  \mean{\sin [n(\psi^{b}_{n}-\psi^{a}_{n})]\sin
    [n(\psi^{d}_{n}-\psi^{c}_{n})]}
  \\&-
  \frac{1}{2}\mean{\cos [n(\psi^{b}_{n}-\psi^{d}_{n})]}
  \mean{\cos [n(\psi^{a}_{n}-\psi^{c}_{n})]}
  \\&+
  \frac{1}{2}\mean{\cos [n(\psi^{a}_{n}-\psi^{d}_{n})]}
  \mean{\cos [n(\psi^{b}_{n}-\psi^{c}_{n})]}. 
\end{split}
\end{equation}
%
Where, $\psi^{i}_{n} \textrm{ (for } i=a,b,c,d)$ are the event-plane angles of $n^{\rm th}$-order harmonic flow. All the three terms contributing to the $T^{obs}_{n}\{ba;dc\}$ observable are calculated independently in this analysis for the Q-level and Particle-level methods.

Since $T_n$ measurements utilizes the event planes reconstructed from four different pseudorapidity regions, ${\Res}(T_{n}\{ba;dc\})$ can be expressed as the product of the four event-plane resolutions,
\begin{equation}\label{resDef}
  {\rm Res}(T_{n}\{ba;dc\})
={\rm Res}(\Psi^{a}_{n}){\Res}(\Psi^{b}_{n}){\Res}(\Psi^{c}_{n}){\Res}(\Psi^{d}_{n}).
\end{equation}
In practice, we estimate ${\Res}(T_{n}\{ba;dc\})$ using formula
\begin{equation}\label{resEq-ql}
\begin{split}
    {\Res}(&T_{n}\{ba;dc\})\\&=\langle\cos [n(\psi^{b}_{n}-\psi^{a}_{n})]\rangle \langle\cos [n(\psi^{d}_{n}-\psi^{c}_{n})]\rangle,
\end{split}
\end{equation}
and identifying the following correlations
\begin{eqnarray}
    \langle\cos [n(\psi^{b}_{n}-\psi^{a}_{n})]\rangle&={\rm Res}(\Psi^{a}_{n}){\Res}(\Psi^{b}_{n}),\label{abRes}\\
    \langle\cos [n(\psi^{d}_{n}-\psi^{c}_{n})]\rangle&={\Res}(\Psi^{c}_{n}){\Res}(\Psi^{d}_{n}),\label{cdRes}
\end{eqnarray}
as the product of the two event-plane resolutions. 
Using Eqs.~\ref{abRes} and~\ref{cdRes}, the product of the four event-plane resolutions in Eq.~\ref{resDef} follows Eq.~\ref{resEq-ql}. 
The calculations of the event-plane resolution from correlation of two event planes  (or three event planes in three sub-event method) is affected by the flow-plane decorrelations.
This effect is relatively small, of the order of 5--15\% based on the previous and our measurements of the ratio correlator (see e.g. Fig.~\ref{fig:r2r3} below), and  significantly smaller than other uncertainties in the measurements. Note that accounting for this effect would slightly {decrease} the values of $T_n$ reported below.

In the Particle-level calculations, the $T_n^{obs}$ correlator and the event-plane resolution correction
${\Res}(T^{p-level}_{n}\{ba;dc\})$ are calculated using azimuthal angles ($\phi^{b}$, $\phi^{c}$) of tracks in mid-pseudorapidity windows $b$ and $c$ instead of $\psi^{b}_{n}$ and
$\psi^{c}_{n}$, 
i.e.,~using formulae
\begin{equation}
  \label{Tn-pl}
\begin{split}
  T_{n}\{ba;dc&\}(\mathrm{Particle\,\text{-}\,level}) \\
  &=\frac{(T_{n}^{obs}\{ba;dc\})^{p-level}}{\Res {\it (T^{p-level}_{n}\{ba;dc\})}},
\end{split}
\end{equation}
\begin{equation}\label{T2explictpl}
\begin{split}
  &(T^{obs}_{n}\{ba;dc\})^{p-level}
  \\&\equiv
  \langle\langle\sin [n(\phi^{b}-\psi^{a}_{n})]
  \sin [n(\psi^{d}_{n}-\phi^{c})]\rangle\rangle
  \\&=
  \mean{\sin [n(\phi^{b}-\psi^{a}_{n})]\sin
    [n(\psi^{d}_{n}-\phi^{c})]}
  \\&-
  \frac{1}{2}\mean{\cos [n(\phi^{b}-\psi^{d}_{n})]}
  \mean{\cos [n(\psi^{a}_{n}-\phi^{c})]}
  \\&+
  \frac{1}{2}\mean{\cos [n(\psi^{a}_{n}-\psi^{d}_{n})]}
  \mean{\cos [n(\phi^{b}-\phi^{c})]}, 
\end{split}
\end{equation}
and
\begin{equation}\label{resEq-pl}
    \begin{split}
      {\Res}(&T^{p-level}_{n}\{ba;dc\})
      \\&=
      \langle\cos [n(\phi^{b}-\psi^{a}_{n})]\rangle \langle\cos [n(\psi_{n}^{d}-\phi^{c})]\rangle
       \\&=
       v_{n}(\eta_{b})v_{n}(\eta_{c}){\Res}(\Psi^{a}_{n}){\Res}(\Psi^{d}_{n}),
    \end{split}
\end{equation}
where $v_{n}(\eta_{b})$ and $v_{n}(\eta_{c})$ represent the $n^{th}$-order flow coefficients such that 
\begin{eqnarray}
    \langle\cos [n(\phi^{b}-\psi^{a}_{n})]\rangle&=v_{n}(\eta_{b}){\Res}(\Psi^{a}_{n}),\label{abFlow}\\
    \langle\cos [n(\phi^{c}-\psi^{d}_{n})]\rangle&=v_{n}(\eta_{c}){\Res}(\Psi^{d}_{n}).\label{cdFlow}
\end{eqnarray} 

Figures.~\ref{fig:t2badc-isobar-res} and~\ref{fig:t3badc-isobar-res} show the centrality dependence of ${\Res}(T_{n}\{ba;dc\})$ and ${\Res}(T^{p-level}_{n}\{ba;dc\})$ for the elliptic and triangular flow in isobar collisions.

The $T^{obs}_{n}\{ba;dc\}$ correlator consists of three terms, and the standard error propagation method is used to calculate the statistical uncertainties on the final observable. 
The uncertainties on the second and third terms in Eqs.~\ref{T2explict} and~\ref{T2explictpl} are negligible compared to that on the first term (see Figs.~\ref{fig:t2badc-isobar-all-comp} and~\ref{fig:t3badc-isobar-all-comp}). 
Hence, the uncertainties on the final observable comes predominantly from the first term even though the three terms in Eqs.~\ref{T2explict} and~\ref{T2explictpl} are correlated. 

Figures~\ref{fig:t2badc-isobar-all-comp} and~\ref{fig:t3badc-isobar-all-comp} show the centrality dependence of different terms contributing to $T^{obs}_{n}\{ba;dc\}$ for isobar collisions.  
Results from the Q-level and Particle-level calculations  are presented side-by-side for better comparison. 
The centrality dependence of different terms in two approaches are slightly different. 
This difference is similar to the one responsible for the difference in the centrality dependence of the event-plane resolution and elliptic flow.

\begin{figure*}[hbtp]
\par\medskip
\centerline{
\includegraphics[angle=0, width=1.0\columnwidth]{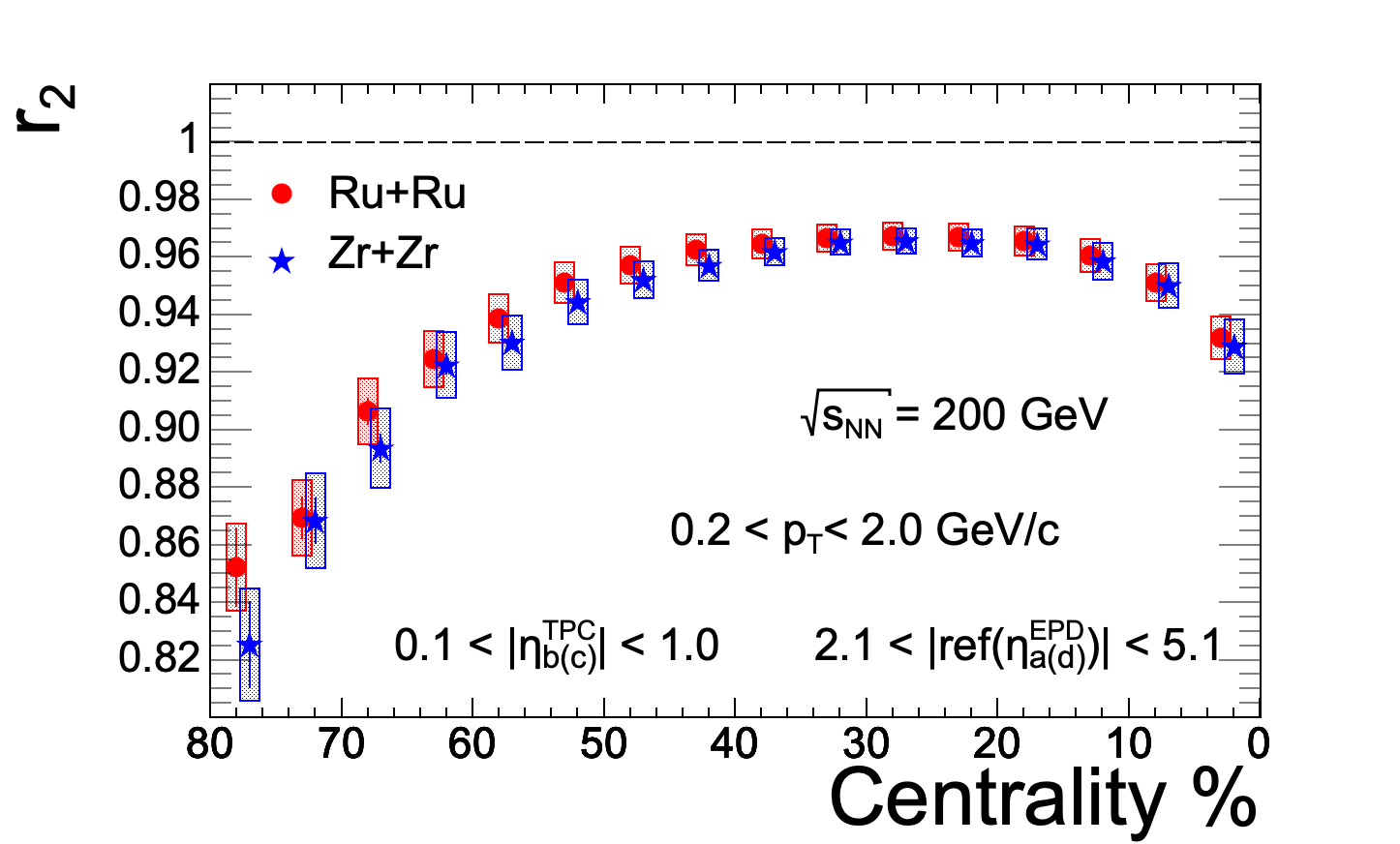}
\includegraphics[width=1.0\columnwidth]{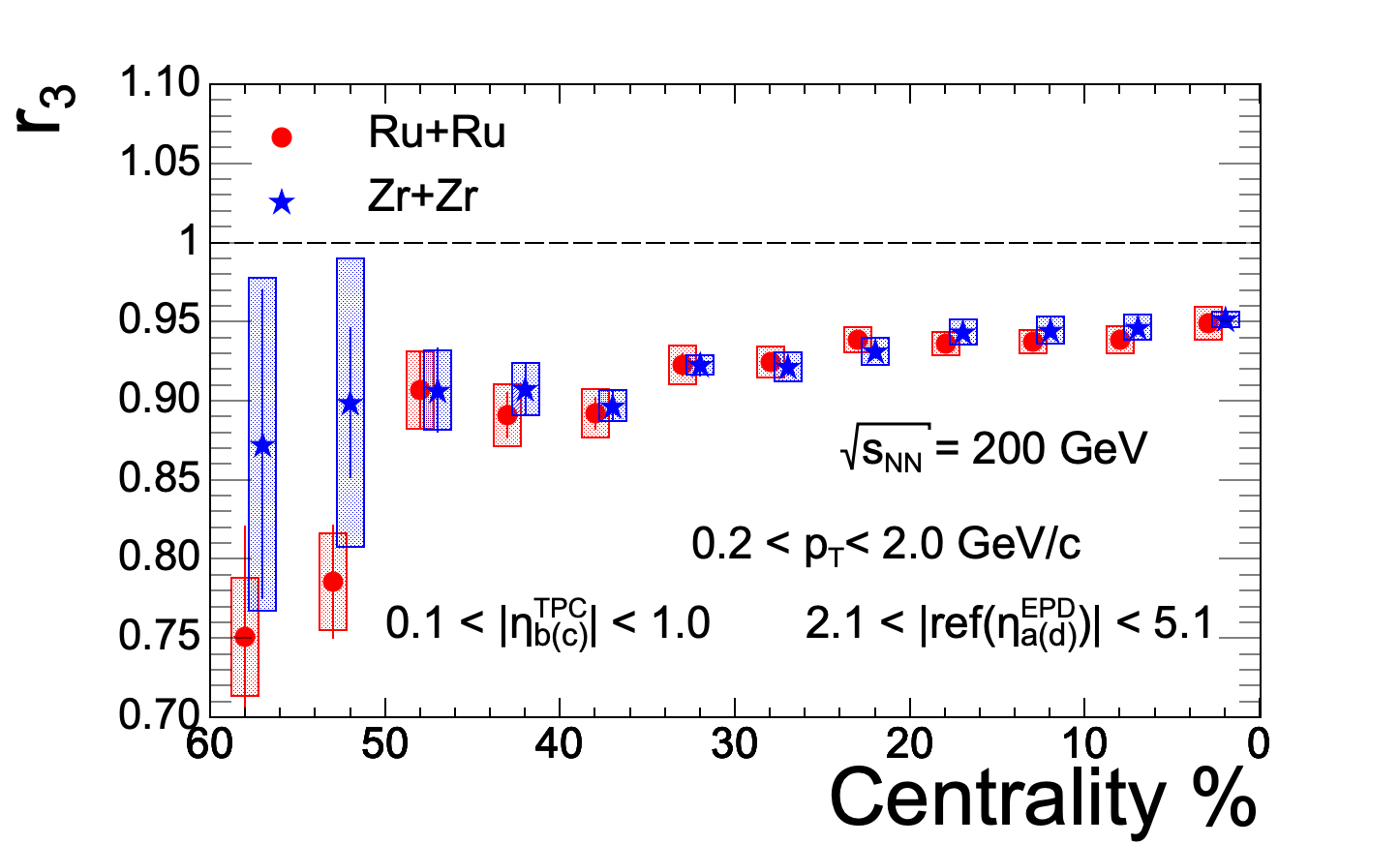}}
\caption{
(Color online) Ratio observable $r_{n}$ (Eq.~\ref{avg_ratio}) as a function of centrality for the second- (left) and third-harmonic (right) anisotropic flow. 
Red circles and blue stars correspond to Ru$+$Ru and Zr$+$Zr collisions, respectively. 
The systematic uncertainties are indicated by boxes. The Zr+Zr data points are shifted towards right for clarity.
}
\label{fig:r2r3}
\end{figure*}

\subsection{Systematic Uncertainty}

Variations of the analysis cuts such as in the $v_{z}$ selection, track DCA, number of reconstructed TPC hits (nHitsFit), and the TPC $\eta$-gap are used for an estimate of the total systematic uncertainty in the measurements. 
All individual contributions to the total systematic uncertainty have been calculated in each centrality bin of $5\%$ step. 
A summary of the nominal and altered cut selections that are used to estimate the systematic uncertainty is provided below.

\begin{itemize}
    \item Nominal cut selection: $-35 < v_{z} < 25$ cm, DCA $ < 3$ cm, nHitsFit $> 15$, $0.2 < p_{\mathrm{T}}< 2.0$ GeV/$c$, $\eta -$gap for the TPC flow vectors reconstruction, $\Delta\eta\text{(TPC)}>0.2$. 
    \item Altered cut selection 1: $-35 < v_{z} < 0$ cm (all other cuts are the same as in the nominal setup).
    \item Altered cut selection 2: DCA $ < 2$ cm (all other cuts are the same as in the nominal setup).
    \item Altered cut selection 3: nHitsFit $ > 20$ (all other cuts are the same as in the nominal setup).
    \item Altered cut selection 4: $\Delta\eta\textrm{(TPC)} > 0.4$ (all other cuts are the same as in the nominal setup).
\end{itemize}

The overall systematic uncertainty is then calculated by adding the individual contributions from different sources in quadrature assuming that the individual systematic uncertainties are uncorrelated. 
The Barlow approach~\cite{barlow2002systematicerrorsfactsfictions} is implemented disentangle statistical fluctuations from systematic uncertainties.
The overall procedure is summarized below:
\begin{enumerate}
  \item 
  Calculate the difference in the observable $Y_{n}$ (where $Y_{n}$ represents either $r_{n}$, $r_{n}^{\Psi}$, or $T_{n}$; see Secs.~\ref{intro-tn} and~\ref{resultSection} for definitions) when computed using the nominal cut and each of the altered cuts: $\Delta
    Y^{i}=Y^{\text{alt},i}_{n}-Y^{\text{nom}}_{n}$.
  \item 
  Calculate the effect of statistical fluctuation defined as
    $\Delta E_{i}=\sqrt{|E_{\text{alt},i}^{2}-E_{\text{nom}}^{2}|}$, where $E$ is the statistical uncertainty of
    $Y_{n}$ measurement.
    The minus sign in the expression for $\Delta E_i$ is used because in two data sets, nominal and altered, one is a subset of another. 
  \item 
  Calculate the systematic uncertainty contribution from each source as $\sigma_{i}=\sqrt{{\Delta Y^{i}}^{2}-{\Delta E_{i}}^{2}}$, but set $\sigma_{i}=0$ if $|\Delta Y^{i}|<\Delta E_{i}$, i.e., if the change $|\Delta Y^{i}|$ is consistent with statistical fluctuations. 
  \item 
  Estimate the total systematic uncertainty by adding the individual contributions from different sources in quadrature using formula $$\sigma_{\text{tot}}=\sqrt{\sum^{N}_{i=1}\sigma_{i}^{2}},$$ where $N=4$, corresponding to four measurements from different cut variations. 
\end{enumerate}

Finally, if the systematic uncertainty estimates obtained using the procedure described above fluctuate across centrality bins, a polynomial fit is applied to smooth the total systematic uncertainty as a function of centrality.

\section{Results}
\label{resultSection}

\subsection{Ratio Observables}

Calculation of the ratio observable involves measuring anisotropic flow over different regions of interest with respect to a reference window. 
We calculated the observable $r_{n}(\eta_{a\rightarrow c/b})$ defined in Eq.~\ref{eq:ratio} for $n= 2$ and $3$ by measuring the ratio of the anisotropic flow in the TPC-west to that in the TPC-east with respect to the fixed reference pseudorapidity window EPD-east. 
\begin{figure*}[hbtp]
\par\medskip
\centerline{\includegraphics[angle=0, width=1.0\columnwidth]{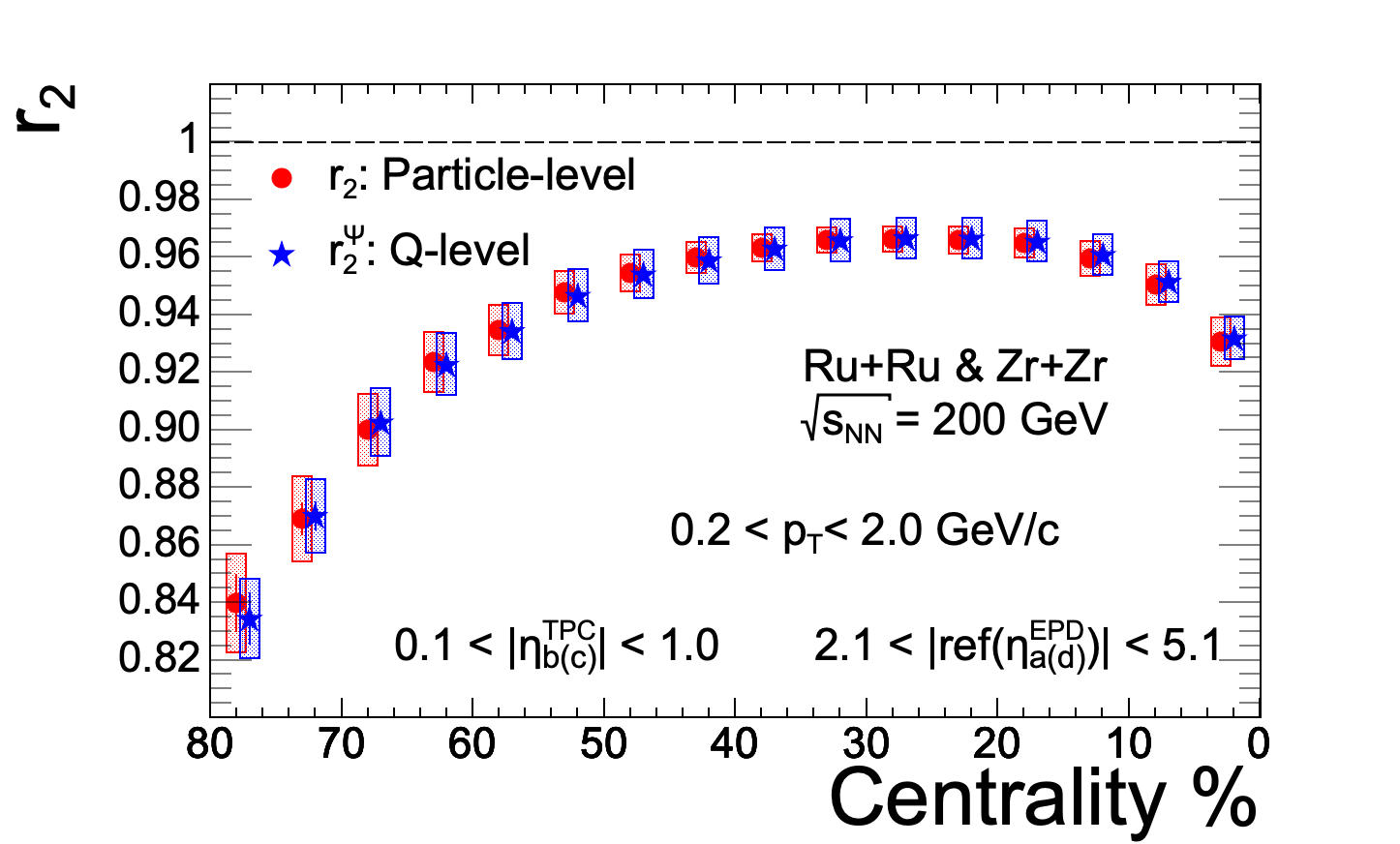}\includegraphics[width=1.0\columnwidth]{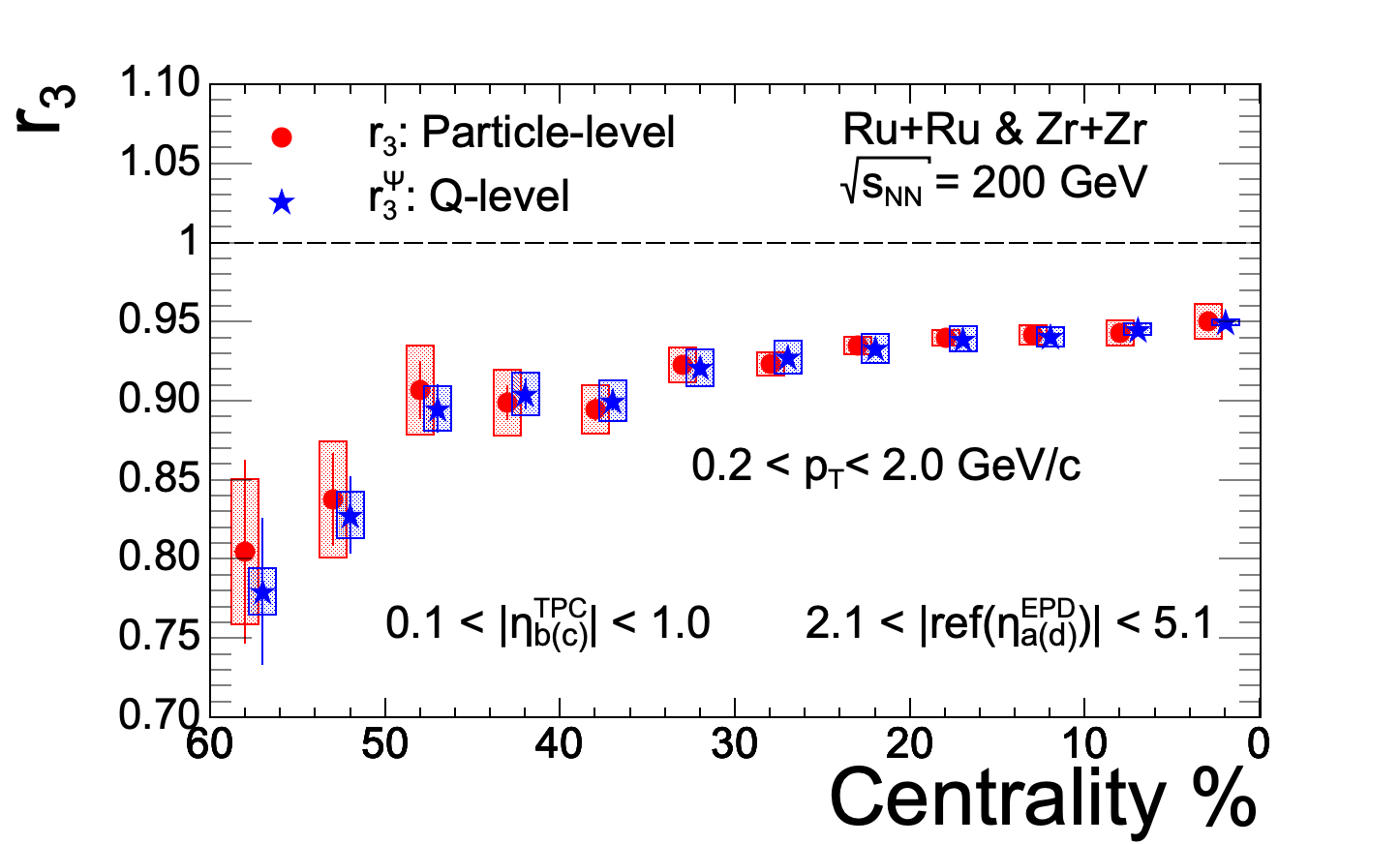}}
\caption{Comparison of ratio observables $r_{n}$ and $r_{n}^{\Psi}$ (defined in Eqs.~\ref{avg_ratio} and~\ref{avg_ratio_ql}) as a function of centrality for the second-- (left) and third-- (right) harmonic anisotropic flow in Ru$+$Ru and Zr$+$Zr collisions. The systematic uncertainties are indicated by boxes. (Color online) Red circles correspond to the Particle-level calculation, and blue stars (shifted horizontally for clarity) correspond to the Q-level calculations (see text for details).
    }
\label{fig:r2r3-ql}
\end{figure*}

The similar ratio $r_{n}(\eta_{d\rightarrow b/c})$ with the EPD-west as a reference pseudorapidity window is calculated using,
\begin{eqnarray}\label{ratio2}
    r_{n}(\eta_{d\rightarrow b/c}) = \frac{\langle\cos[n(\phi^{b}-\psi^{d}_{n})]\rangle}{\langle\cos[n(\phi^{c}-\psi^{d}_{n})]\rangle}. 
\end{eqnarray}

Taking advantage of the symmetric STAR detector system, we combined the two measurements, $r_{n}(\eta_{a\rightarrow c/b})$ and $r_{n}(\eta_{d\rightarrow b/c})$, to improve statistical precision in the final result using,
\begin{eqnarray}\label{avg_ratio}
    r_{n} = \frac{1}{2} [r_{n}(\eta_{a\rightarrow c/b}) + r_{n}(\eta_{d\rightarrow b/c})].
\end{eqnarray}

Additionally, the decorrelations of the flow-plane angles are also calculated using the event-plane angles $\psi_{n}$ in the ratio formulation using,
\begin{equation}
\label{avg_ratio_ql}
\begin{split}
    r^{\Psi}_{n} &= \frac{1}{2} [r^{\Psi}_{n}(\eta_{a\rightarrow c/b}) + r^{\Psi}_{n}(\eta_{d\rightarrow b/c})]\\
    &=\frac{1}{2} \bigg[\frac{\langle\cos[n(\psi^{c}_{n}-\psi^{a}_{n})]\rangle}{\langle\cos[n(\psi^{b}_{n}-\psi^{a}_{n})]\rangle}+\frac{\langle\cos[n(\psi^{b}_{n}-\psi^{d}_{n})]\rangle}{\langle\cos[n(\psi^{c}_{n}-\psi^{d}_{n})]\rangle}\bigg].
\end{split}
\end{equation}

The measurements of $r_{n}$ and $r^{\Psi}_{n}$ correspond to the factorization ratios obtained through the {Particle-level} and {Q-level} methods discussed earlier. 
Note that the Q-level calculation utilizes $\psi_{n}$'s from the TPC, reconstructed using $p_{\mathrm{T}}$ weights, resulting in better event-plane resolutions.

Figure~\ref{fig:r2r3} presents the results for the factorization ratio $r_{n}$ as a function of centrality for the second and third harmonic flow in isobar collisions. 
The $r_3$ results for $60-80\%$ centrality range have large uncertainties due to low multiplicity and are not shown. In mid-peripheral collisions, small differences in the $r_{n}$ measurements are observed between the Ru+Ru and Zr+Zr collision systems.

The ratios $r_{n}$ and $r^{\Psi}_{n}$ are less than unity as expected due to decorrelations increasing further away from the reference pseudorapidity window.  All these ratio measurements are consistent within the statistical uncertainties for the Ru+Ru and
Zr+Zr isobar collisions. 
Additionally, the $r_{n}$ and $r^{\Psi}_{n}$ results are found to be in good agreement with each other within experimental uncertainties at all centralities despite our use of flow vectors reconstructed using different weights in the two measurements. Therefore, the comparison plot of $r_{n}$ and $r^{\Psi}_{n}$ as shown in Fig.~\ref{fig:r2r3-ql} is presented only for the combined isobar data set rather than for the individual Ru+Ru and Zr+Zr data sets.

\begin{figure*}[htbp!]
\centerline{\includegraphics[angle=0, width=1.0\columnwidth]{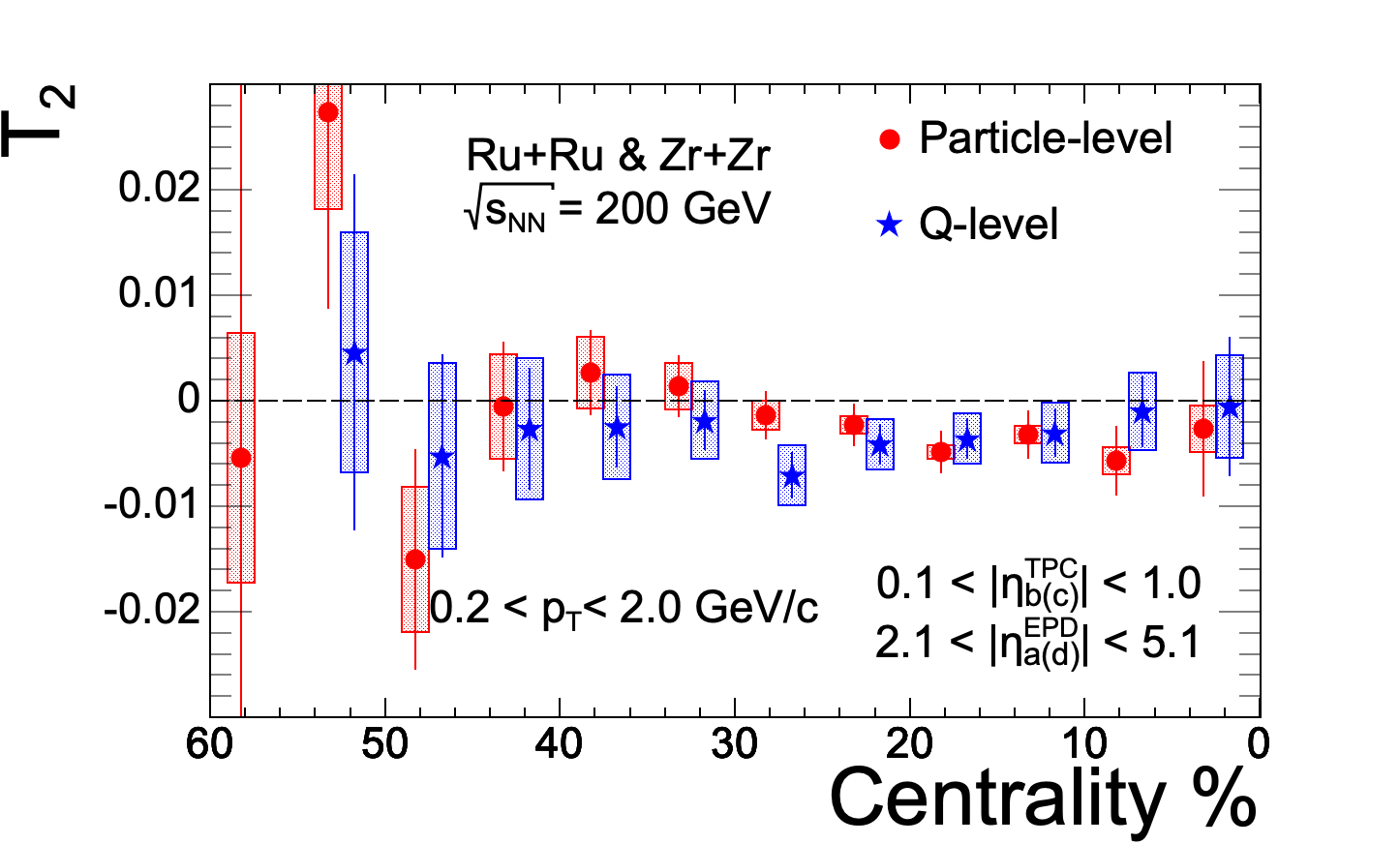}\includegraphics[width=1.0\columnwidth]{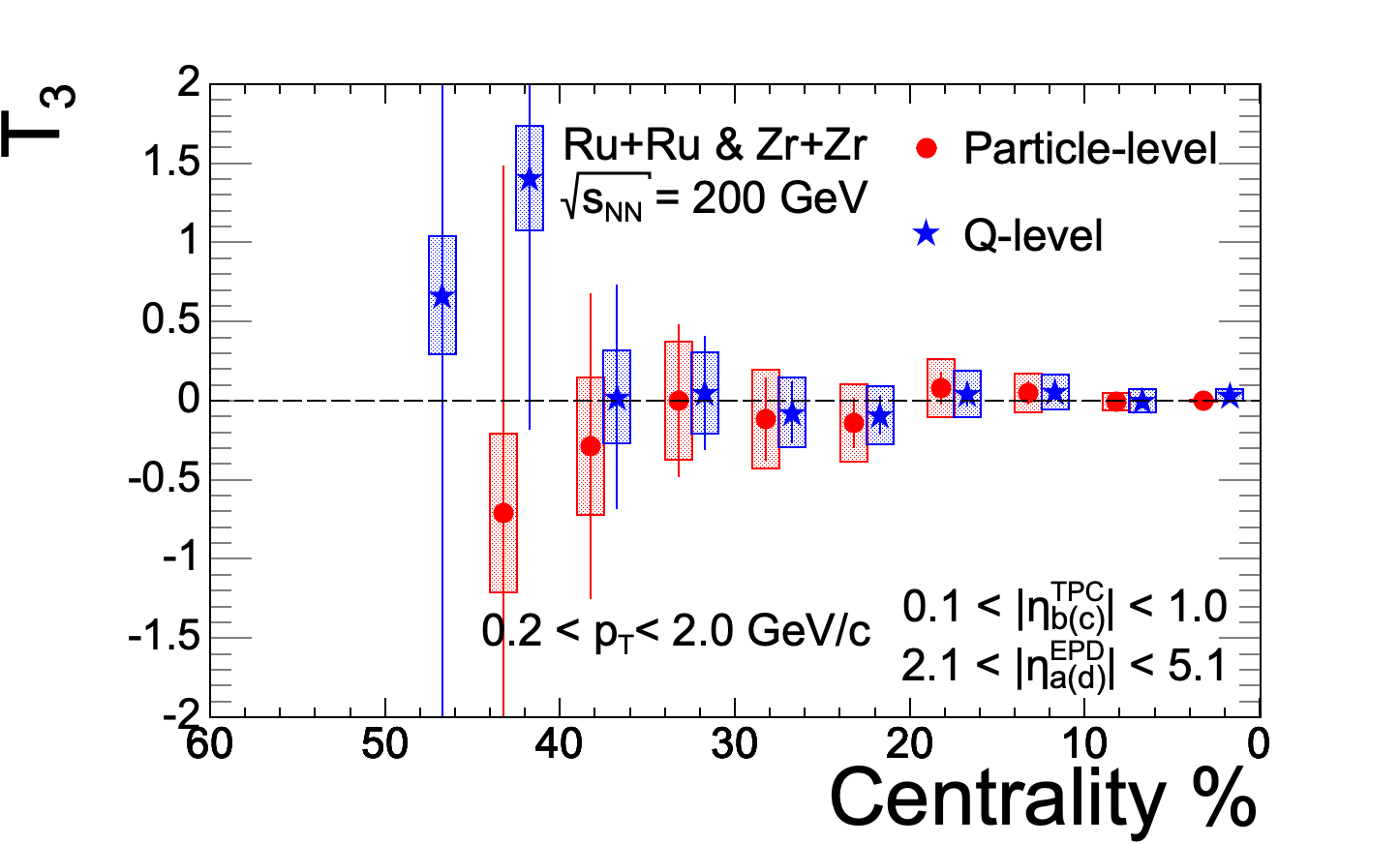}}
\caption{(Color online) $T_{n}\{ba;dc\}$ observable 
as a function of centrality for the elliptic (left) and
  triangular (right) flow in Ru+Ru
  and Zr+Zr collisions at $\sqrt{s_{_\mathrm{NN}}}=200$ GeV. 
  The systematic uncertainties are indicated by boxes. 
  Red circles correspond to the Particle-level calculation, and blue stars (shifted horizontally for clarity) correspond to the Q-level
  calculations.
    }
\label{fig:tnbadc-isobar}
\end{figure*}

The centrality dependence of $r_{n}$ qualitatively can be explained as follows. 
In the mid-central collision, the flow is large and the effect of fluctuations is relatively small, resulting in the $r_2$ values closer to unity. 
On the other hand, in central collisions the flow is small and thus the relative effect of fluctuations increases. In peripheral (low multiplicity) collisions the flow is decreasing while the fluctuations become larger due to the small number of participants. 
The $r_3$ is the largest in most central collisions where the triangular flow becomes the strongest. Similar reasoning applies to explain the feature of $r^{\Psi}_{n}$ as a function of centrality.

Figure~\ref{fig:r2r3-ql} shows comparison between $r^{\Psi}_{n}$ (Eq.~\ref{avg_ratio_ql}, Q-level) and $r_{n}$ (Eq.~\ref{avg_ratio}, Particle-level) measurements for the second- and third-order flow in isobar collisions. 
The combined dataset of Ru+Ru and Zr+Zr collisions is used to improve the statistical precision. 
Using $\Psi_{n}$'s with improved resolutions due to $p_{\mathrm{T}}$ weighting, one might intuitively anticipate relatively smaller decorrelation in the Q-level measurements. 
However, Fig.~\ref{fig:r2r3-ql} shows a negligible difference between the Q-level and Particle-level ratio measurements in each centrality bin, despite the use of $p_{\mathrm{T}}$ weights in the former. 
This indicates that our ratio measurements primarily reflect the dynamic effects, including contributions from both the random-walk and the long-range systematic variations of the flow plane, as discussed in Sec.~\ref{intro}. 
Such variations are believed to be stem out from the initial state fluctuations of the collision systems.

\subsection{\texorpdfstring{$T_{n}$}{Lg} Observables}

Previous subsections reported the measurements of observed $T^{obs}_{n}\{ba;dc\}$ and corresponding event-plane resolution correction terms for the elliptic and triangular flow for isobar collisions. 
The final results for the $T_{n}$ observable, corrected for the event-plane resolution, are discussed in this section. 
Regardless of the small difference~\cite{Xu:2017zcn} in the nuclear deformation of the two isobar species, the results shown in previous sections are consistent within the statistical uncertainties for Ru+Ru and Zr+Zr collisions. 
Hence, the final results are calculated using the combined Ru+Ru and Zr+Zr collision data for a better statistical precision over the wider range of centrality.
\begin{figure}[htbp!]
     \centerline{\includegraphics[width =\columnwidth]{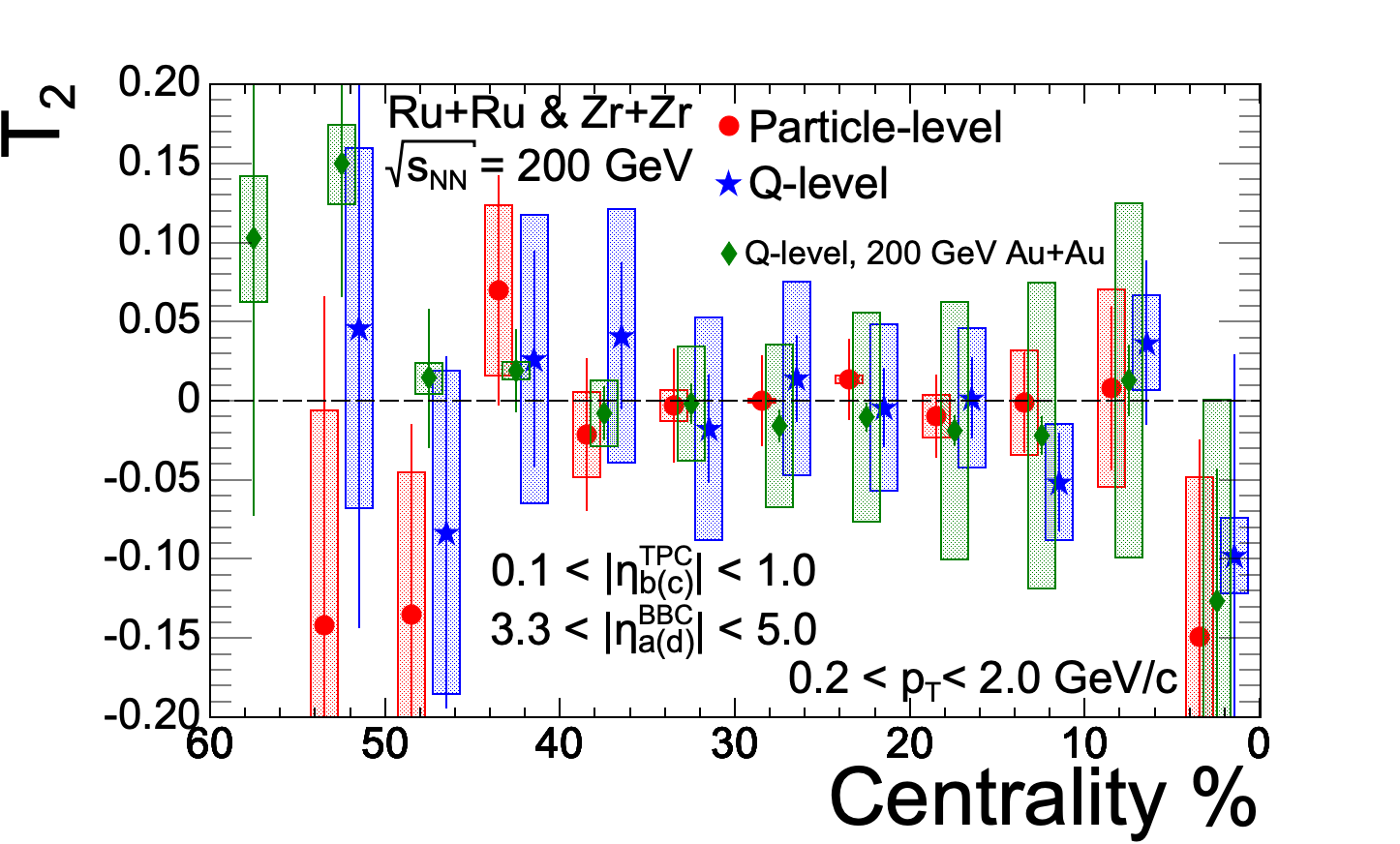}}
\caption {Same as Fig.~\ref{fig:tnbadc-isobar} but using forward and backward flow planes from BBC instead of EPD  for isobar  and Au+Au collisions at 
  $\sqrt{s_{_{\mathrm{NN}}}}=200$~GeV. 
  Green diamond data points correspond to the Q-level calculations in Au+Au collisions.
}
\label{fig:tnfinal-tpcbbc}
\end{figure}

Figures~\ref{fig:tnbadc-isobar} and~\ref{fig:tnfinal-tpcbbc} present the results of $T_{n}\{ba;dc\}$ measurements as a function of centrality for the elliptic and triangular flow in isobar collisions at $\sqrt{s_{_{\mathrm{NN}}}}=200$~GeV. The results for the $60-80\%$ centrality range have large uncertainties due to low multiplicities and are not shown. The comparison between the results obtained using the Q-level and Particle-level approaches has been shown for a centrality range of $0-60\%$ in $12$ centrality bins. The results from the two approaches are in good agreement with each other within experimental uncertainties. 
The $T_{2}$ (Q-level) measurement averaged over the $10-30\%$ centrality range is $-0.004 \pm 0.001 (\rm stat) \pm0.002(\rm syst)$ and consistent with zero, within experimental uncertainties, for other centralities. 
The $T_{3}$ result is consistent with zero within experimental uncertainties in all centrality bins. Such small values of $T_n$ observables from experimental data support a ``random-walk'' variation of flow-plane angles with pseudorapidity (see Sec.~\ref{discussionSummay} for detailed discussion and summary).

The similar measurements for the elliptic flow in Au+Au and isobar collisions at $\sqrt{s_{_{\mathrm{NN}}}}=200$ GeV using four flow planes reconstructed from the positive and negative pseudorapidity sides of the TPC, and BBC are shown in Fig.~\ref{fig:tnfinal-tpcbbc}. 
Within the statistical uncertainty, the $T_{n}$ measurements in Au+Au and isobar collisions are found to be compatible with each other indicating weak dependence of the results on the system size.

Our $T_{n}$ measurements are in agreement, within the statistical and systematic uncertainties,  with the AMPT model calculations from Ref.~\cite{Xu:2020koy} for Au+Au collisions at $\sqrt{s_{_{\mathrm{NN}}}}=200$~GeV. 
The calculation reported in Ref.~\cite{Xu:2020koy} yields slightly positive $T_{2}$ values, approximately on the order of $10^{-3}$ for events within the $20-40\%$ centrality range. 
Due to differences in the ordering of the flow-plane angles, a positive $T_{2}$ as defined in Ref.~\cite{Xu:2020koy} corresponds to a negative $T_{2}$ in our convention.

\section{Discussions and summary}
\label{discussionSummay}

We measured the four-plane correlator $T_{n}$, and the factorization ratios $r_{n}$ and $r_{n}^{\Psi}$, in isobar (Ru+Ru and Zr+Zr) and Au+Au collisions at $\sqrt{s_{_{\mathrm{NN}}}}=200$~GeV. 
These measurements focus on $n=2$ and $n=3$ flow harmonics, testing the longitudinal flow-plane decorrelation phenomena in heavy-ion collisions.

In mid-central collisions ($10-30\%$ centrality range), $T_{2}\{ba;dc\}= -0.004 \pm 0.001 (\mathrm{ stat})\pm0.002(\mathrm{ syst})$ have been observed independent of the collision system and consistent with zero within experimental uncertainties at other centralities. 
The $T_{3}\{ba;dc\}$ result is consistent with zero within experimental uncertainties at all centrality bins. 
The factorization ratio observables $r^{\Psi}_{n}$ and $r_{n}$ are found to be similar within experimental uncertainties across all centralities, despite the use of $p_{T}$ weights in the former. 
The $r_{2}$ observable for $n=2$ flow harmonics reaches its maximum value of $r_2\approx 0.96$ in the $10-30\%$ centrality range. 
Similarly, the maximum value of $r_{3}\approx 0.95$ is observed in the most central $0-5\%$ centrality bin. 
As outlined in Sec.~\ref{intro}, such values of the factorization ratio correspond to flow decorrelations of approximately $\frac{n^{2}}{2}\mean{(\Delta\Psi^{b\rightarrow c}_n)^2} \sim 0.04$ and $\sim 0.05$ for $n=2$ and $n=3$ harmonics, respectively. 
Considering these decorrelation estimates, and assuming a similar order of decorrelation between mid- and forward (backward) pseudorapidities, the prediction for the $T_{n}\{ba;dc\}\sim n^{2}\langle \Delta\Psi^{a\rightarrow  b}_{n} \cdot \Delta\Psi^{c\rightarrow d}_{n}\rangle$ observable would be at least $10$ times larger in magnitude than those observed.

Additionally, the measured values of $T_{n}$ are significantly smaller than what one would expect in the scenario of monotonic variation of the flow angles between forward and backward pseudorapidity windows determined by eccentricities calculated using participant nucleons from projectile and target nuclei. 
According to calculations in Ref.~\cite{PhysRevC.90.034915}, the RMS value of the forward-backward twist $(2\Psi_B-2\Psi_F)$ distribution in such an approach is about $0.5$ for mid-central collisions, which would translate to $T_2\approx 0.5^2/4$, again much larger than values observed experimentally.

As noticed in Sec.~\ref{intro}, effective flow magnitude weights likely bias the measurements of $r_n$ and $T_n$ correlators toward smaller values, but this effect would not be able to explain an order of magnitude difference in the above estimates. 
In the future high statistics measurements, it will be possible to study the effect of flow magnitude fluctuations by introducing additional $v_n$ weights in the definitions of Eq.~\ref{T2def}. 

The above considerations lead to a conclusion that our results favor a picture of a ``random-walk'' variation of the flow-plane angles, with a pseudorapidity correlation length smaller than the pseudorapidity region under study. 
In this scenario, the variation of the flow-plane angle along pseudorapidity is dominated by random fluctuations with the correlation length in pseudorapidity being much smaller than the size of pseudorapidity windows used in this analysis.
These type of decorrelations contribute to the ratio observable $r_n$ but not to $T_n$ as the contribution of any low order flow plane correlations are explicitly removed from $T_n$ correlator. 
As was noted in Ref.~\cite{Xu:2020koy}, this type of decorrelation would also lead to almost no dependence of $r_n$ on the reference pseudorapidity, which is consistent with experimental observations.  
Slightly negative result for $T_{n}$ could indicate that both flow planes at mid-pseudorapidity are fluctuating together relative to those at the forward and backward pseudorapidities. 
Our results provide new insights into the decorrelation patterns within the system and offer a quantitative estimate of possible systematic variations in anisotropic flow angles such as the “twist” between forward and backward pseudorapidity regions. 
The measurements open new avenues for exploring the three-dimensional structure and the time evolution of the QGP formed in heavy-ion collisions. 


\begin{acknowledgments}
We thank the RHIC Operations Group and SDCC at BNL, the NERSC Center at LBNL, and the Open Science Grid consortium for providing resources and support.  This work was supported in part by the Office of Nuclear Physics within the U.S. DOE Office of Science, the U.S. National Science Foundation, National Natural Science Foundation of China, Chinese Academy of Science, the Ministry of Science and Technology of China and the Chinese Ministry of Education, NSTC Taipei, the National Research Foundation of Korea, Czech Science Foundation and Ministry of Education, Youth and Sports of the Czech Republic, Hungarian National Research, Development and Innovation Office, New National Excellency Programme of the Hungarian Ministry of Human Capacities, Department of Atomic Energy and Department of Science and Technology of the Government of India, the National Science Centre and WUT ID-UB of Poland, German Bundesministerium f\"ur Bildung, Wissenschaft, Forschung and Technologie (BMBF), Helmholtz Association, Ministry of Education, Culture, Sports, Science, and Technology (MEXT), and Japan Society for the Promotion of Science (JSPS).

\end{acknowledgments}

\bibliography{main.bib}

\end{document}